\documentclass[nofootinbib,aps]{revtex4}

\usepackage{amstext,amsmath,amssymb}
\usepackage[dvips]{graphicx}
\usepackage{latexsym}

\newcommand{\Ref}[1]{(\ref{#1})}

\newcommand{\N}{\mathbb{N}}
\newcommand{\Z}{\mathbb{Z}}

\newcommand{\R}{\mathbb{R}}
\newcommand{\C}{\mathbb{C}}

\def\be{\begin{equation}}
\def\ee{\end{equation}}
\def\bes{\begin{eqnarray}}
\def\ees{\end{eqnarray}}
\def\nn{\nonumber}
\def\arr{\rightarrow}

\def\Om{\Omega}

\def\la{\langle}
\def\ra{\rangle}
\def\f{\frac}

\def\wtl{\widetilde}
\def\what{\widehat}
\def\hL{\what{L}}

\newcommand{\SU}{\mathrm{SU}}
\newcommand{\SO}{\mathrm{SO}}
\def\wT{\what{T}}
\def\pp{\partial}

\newcommand{\lalg}[1]{\mathfrak{#1}}

\newcommand{\mat}[2]{\left(\begin{array}{#1} #2\end{array}\right)}

\newcommand{\bk}[1]{\langle #1 \rangle}

\begin{document}

\title{About Lorentz invariance in a discrete quantum setting}

\author{{\bf Etera R. Livine}\footnote{elivine@perimeterinstitute.ca}}
\affiliation{Perimeter Institute, 35 King Street North
Waterloo, Ontario Canada N2J 2W9}
\author{{\bf Daniele Oriti}\footnote{d.oriti@damtp.cam.ac.uk}}
\affiliation{Department of Applied Mathematics and Theoretical Physics
\\ Centre for Mathematical Sciences, University of Cambridge,
Wilberforce Road, Cambridge CB3 0WA, England, EU \\ and \\ Girton College, University of Cambridge \\ Cambridge CB3 0JG, England, EU}

\begin{abstract}

\begin{center}
{\small ABSTRACT}
\end{center}

A common misconception is that Lorentz invariance is inconsistent
with a discrete spacetime structure and a minimal length: under
Lorentz contraction, a Planck length ruler would be seen as
smaller by a boosted observer. We argue that in the context of
quantum gravity, the distance between two points becomes an
operator and show through a toy model, inspired by Loop Quantum Gravity, that the notion of a quantum
of geometry and of discrete spectra of geometric operators, is not inconsistent with Lorentz invariance. The main
feature of the model is that a state of definite length for a
given observer turns into a superposition of eigenstates of the length operator when seen by a boosted observer. More generally, we discuss the issue of actually measuring
distances taking into account the limitations imposed by quantum
gravity considerations and we analyze the notion of distance and the phenomenon of Lorentz contraction in
the framework of ``deformed (or doubly) special relativity'' (DSR), which
tentatively provides an effective description of quantum gravity
around a flat background. In order to do this we study the Hilbert space structure of DSR, and study various quantum geometric operators acting on it and analyze their spectral properties. We also discuss the notion of spacetime point in DSR in terms of coherent states. We show how the way Lorentz invariance is preserved in this context is analogous to that in the toy model.

\end{abstract}

\maketitle

\section{Introduction}

\begin{flushright}
{\small {\it `` It is usually assumed that space-time is a continuum.
This assumption \\ is not required by Lorentz invariance.
In this paper we give an example \\ of a
Lorentz invariant discrete space-time''} \\
{\bf Snyder, 1947}}
\end{flushright}

Gravitation is geometry and measurements of distances are measurements
of properties of the gravitational field. Geometric quantities are the
observable properties of the
gravitational field. If the gravitational field is quantized, then
this would result in geometric observables being quantum operators of
which it is sensible to compute the spectrum and to find the set of
eigenstates.
A first expected effect of the combination of the principles of
General Relativity and Quantum Mechanics is the appearance of a
fundamental length scale, the Planck length $l_p = \sqrt{\frac{\hbar
    G}{c^3}}$, and of a fundamental time scale, the Planck time
$\tau_p =\sqrt{\frac{\hbar G}{c^5}}$. These are supposed to represent
a lower bound on the measurement of spacelike and timelike distances
in a quantum gravity setting, where both the quantum and dynamical
properties of the geometry are taken into account, and the very notion of
distance is
expected to lose physical meaning below such scales. Also, these new
fundamental constants would act as a natural cutoff for any field
living on a manifold endowed with a quantum geometry \cite{Thiemann}, also
explaining
puzzling features of black hole physics \cite{bekenstein}.

Provisional results confirming these expectations come from both
quantum gravity, in the canonical \cite{ThieRev, CarloRev} as well as in
some
versions of the covariant approach \cite{Daniele, alejandro}, and string
theory
  \cite{AmatiVeneziano}.

Moreover, it is a intriguing possibility that, as it happens for many
classically continuous observables, geometric operators would turn out
to have a discrete spectrum in a fully quantized theory. Geometric
operators such as lengths or areas, or time intervals, would then
assume, when measured on states of the quantum gravitational field,
only a discrete set of possible values, with probabilities computable
within the theory itself. This possibility, again,  is realized in both
Loop
Quantum Gravity and Spin Foam Models.

This, of course, has to be true also if the gravitational field is
in a particular quantum state being a ``Minkowski quantum state'',
in the sense that it approximates a flat Minkowski geometry on
scales larger than the Planck scale, i.e. in a
semiclassical/continuum approximation, provided we do not neglect
completely the quantum nature of geometric observables, by taking
$\hbar \rightarrow 0$.

Needless to say, this is a very different picture of spacetime
geometry with respect to the classical one we are accustomed with,
particularly with respect to the simple picture of spacetime geometry
in a classical flat Minkowski background.

This conceptual jump naturally raises confusion and a host of questions.

In particular, what is the relationship between this lowest bound in
distances, or the possible
discreteness of the spectrum of lengths, and the usual continuous
symmetries of classical GR or special relativity. We know that, in a
flat classical geometry, a
spacelike length at rest with respect to a given observer will result
in being contracted when measured by a second observer related to the
first by a Lorentz transformation (boost), while time intervals are
dilated by the same kind of transformation. The contraction and
dilatation are moreover governed by a continuous parameter, the boost
rapidity. It seems then that both the idea of a minimal length and
that of a discrete set of possible values for lengths and time
intervals are at odds with what we know about Lorentz symmetry.

If an observer has verified the existence of a minimal length being
the Planck one for a given object, a second boosted observer will see a
smaller length for the
same object, as a result of Lorentz contraction. If the first observer
has verified the existence of a discrete spectrum for the length of
that object, the second boosted observer will fail to detect this
discreteness because of the continuous nature of the Lorentz
symmetry. In any case, it seems that neither the lowest Planckian bound
for
distances, nor their spectrum can possibly be an invariant property of
the theory, if we require it to be Lorentz symmetric in the flat regime.

Of course none of these observations is conclusive, and the problem as
explained above is pretty naively posed, since it pretends to apply
what we know about Lorentz symmetry in flat Minkowski spacetime to a
fully quantum geometric theory in which a breakdown of all
conventional notions of smooth geometry are likely to take place.

Nevertheless, any quantum
gravity model predicting either discrete spectra of geometric operators
or minimal and non-zero eigenvalues for them (or both) should face
these issues and show how the apparent paradox is solved.

Does a quantum gravity theory with an invariant length and a discrete
spectrum for geometric observables necessarily break Lorentz symmetry
or necessarily require some sort of modification/deformation of it?
The answer, as we will see, is simply ``no''.

In this paper, we tackle this issue of the action of length
contraction/dilatation in a quantum gravity setting in a simplified toy
model inspired
by loop quantum gravity, and then in the context of Deformed (or Doubly) Special
Relativity, considered as an effective description of some features of
a full quantum gravitational theory.

\subsection{Loop Quantum Gravity discreteness}

Loop Quantum Gravity \cite{ThieRev, CarloRev} is a well-developed theory
for dealing with the quantum properties
of geometry, and as such all these issues are to be addressed in it.
Indeed, in the Loop Quantum Gravity context we see at work both the
features we anticipated above: it predicts a precise spectrum for
geometrical
quantities, such as lengths, areas and volumes, and in many cases
these spectra turn out to be discrete, and result in having a lowest
bound at the Planck scale.  Here we focus on what happens
to lengths, and therefore we analyze the situation in $2+1$ gravity,
where lengths operators are most easily analyzed when the theory is
quantized in the loop approach. An additional reason for dealing with
$2+1$ gravity is that the theory (in the absence of a cosmological
constant) is flat, and so the implementation of Lorentz symmetry is most
easily analyzed. Of course, we face the same conceptual/technical
issues as outlined above.

In particular, in the Lorentzian version of $2+1$ Loop Quantum
Gravity we have a prediction of a continuous spectrum for
spacelike distances and of a discrete spectrum for timelike
intervals \cite{lqg3d}. This is also fully consistent with the
picture obtained from the corresponding covariant approach, i.e.
from Lorentzian spin foam models in 3d \cite{Laurent}. Indeed, in
this approach the length operator acts diagonally on spin networks
whose edges are labeled by representations of $\SO(2,1)$.
one edge
As there are two types of (principal) representations of
$\SO(2,1)$, the spectrum of the length operator is
\be
\hat{L}\,\Psi\,=\,l_p\,\sqrt{\rho^2\,+\,\frac{1}{4}}\,\Psi
\qquad \textrm{and} \qquad
\hat{L}\,\Psi\,=\,i\tau_p\,\epsilon\,\sqrt{-n(n-1)}\,\Psi,
\ee
where $\rho\in\R_+$ or $(\epsilon= \pm 1, n\in
\mathbb{N})$ label the (unitary) representations of $\SO(2,1)$.

Let us point out that there exists a quantization ambiguity
resulting from the regularization procedure. This leads to an
alternative length spectrum:
\be
\hat{L}_s\,\Psi\,=\,l_p\,\rho\,\Psi
\qquad \textrm{and} \qquad
\hat{L}\,\Psi\,=\,\tau_p\,\epsilon\,i\,(n\,-\,\frac{1}{2})\,\Psi
\ee

The length spectrum with continuous representations used has the
interpretation of a quantization of spacelike distances, it is
continuous and, at least in the first case, presents a lowest bound on
allowed values corresponding to (half) the Planck length. When discrete
representations are used we have a quantization of timelike distances
(intervals), the spectrum is discrete and, at least in the second case,
it presents a lowest bound on the allowed values, equal again to
(half) the Planck time\footnotemark .

\footnotetext{These are the formal results of the theory regarding the
  spectrum of distances, and refer to observables which can be in
  principle measured by suitably planned (and precise enough) experiments.
However, much
  care should be used in interpreting them, and in relating them to
  actual observations. The question is: what is really measurable
  (experimentally)? The issue is what a correct operational definition
  of a spacelike or timelike distance is, what would be a concrete and
  realistic procedure for measuring them. Consider for example that in
special
  relativistic settings the measurement of spacelike distances is
  achieved through emission and subsequent reception of light signals
  and measurement of timelike intervals giving information about the
  traveled distance by the light. If anything similar is true also in
  a quantum gravity context, then a discreteness of
measured timelike intervals may imply a discreteness of measured spacelike
distances, when an appropriate operational definition of measurement
is given}.

Of course, the above spectra are for the length of spacetime
intervals in flat space (because 3d gravity admits only flat
solutions), and these are not supposed to change under boosts, which
are in fact isometries of the flat metric. Therefore obtaining a
discrete spectrum like the one given above is not at odds with Lorentz
invariance. However it suggests that other geometric quantities, that
are not invariant under boosts, and indeed depend on a choice of
observer, like time intervals or space distances, may have a discrete
spectrum as well, and the question is whether such quantities maintain
their discrete features or not under boosts. The analogy is with the
usual coordinate representation of the spacetime infinitesimal
interval in Minkowski space: $ds^2 = - dt^2 + dx^2 + dy^2$ , in a given
coordinate frame, so for a given observer; its integral
between two events $A$ and $B$ gives their invariant spacetime
distance (square) $L_{AB}^2$ that is the result of a time interval
$T_{AB}^2$ and a space distance $S_{AB}^2$, as $L_{AB}^2 = - T^2_{AB} +
S_{AB}^2$; of course, neither of the last two quantities is observer
independent and indeed it is these two that change under boosts,
i.e. in going to another observer (coordinate choice), but leaving the
full spacetime length invariant. But if $L_{AB}^2$ is discrete, then
it is likely that $T_{AB}^2$ and  $S_{AB}^2$ can be discrete as well;
the question is what is their behavior under boosts. This analogy
will be exploited in constructing a simplified model of an operator
representing time
interval measurements in the following.

We are then faced with the following theoretical/mathematical issue: is it
possible to have a consistent
length contraction or time dilatation for boosted observers in the
presence of the above (discrete) length spectrum? Assuming that an
inertial observer measures a
quantum length with results governed by a spectrum of the kind given
above, what would
a boosted observer measure for the same object? In other words, how
does a Lorentz transformation change the previous equation?

There are two possibilities. It is possible that the spectrum will
change, meaning that it does not represent a Lorentz invariant
property of the theory; this may happen if to a boosted observer it
corresponds a change of the representation of the operators defining
the quantum theory, giving an inequivalent quantization; in the
context of the $\SU(2)$-based Loop Quantum Gravity this would be for
example a
change of the Immirzi parameter. The second possibility is to have an
invariant spectrum on which both the initial and the boosted observer
agree, although they may disagree on the expression for both the
length operator and the quantum state of the gravitational field. In
this situation:
\be
\hat{L}_{boosted}(\beta)\, =\,
U(\beta)\,\hat{L}\,U^{-1}(\beta)\;\;\;\;\;\;\;\;\;\;\;{\text
  and}\;\;\;\;\Psi_{boosted}(\beta)\,=\,U(\beta)\,\Psi
\ee
where $U(\beta)$ is the operator corresponding to a Lorentz boost in
an appropriate representation of the Lorentz group, but still
\be
\hat{L}_{boosted}\,\Psi_{boosted}\,=\,L\,\Psi_{boosted},
\ee
The usual Lorentz contraction/dilatation will be found for the
mean value of the operator in any given state. Of course this is
possible only if $U$ is in a unitary representation of the Lorentz
group (remember that the only representations labeling the edges
of spin networks in Loop Quantum Gravity are the unitary ones) and
if the boosted and un-boosted length operators do not commute,
$\left[\hat{L},\hat{L}_{boosted}\right]\,\neq 0$. This possibility
was indeed proposed, as a way to solve the apparent paradox, and
investigated in \cite{simone}.

This second resolution of the naive paradox, leading to an Lorentz
invariant spectrum of quantum geometric operators, is not at all a
result of pure wishful thinking or a lucky but unlikely to happen
possibility. It is nothing else than what happens in ordinary
quantum mechanics as a natural resolution of an analogous apparent
paradox, in which the theory predicts a discrete spectrum with
lowest bound for an observable that is transformed by a continuous
transformation when different observers are considered: angular
momentum. The angular momentum of a particle may be quantized and
its spectrum is discrete, with its lowest possible eigenvalue
being $1/2$ for a fermionic particle. Rotations are continuous
transformations and are represented in the quantum theory by
quantum operators labeled by a continuous angular parameter
$\theta$. However, the spectrum of the angular momentum is
nevertheless a rotational invariant property of the theory, in the
sense that two observers related by a rotation one to the other
will measure a different operator on a different state of the
particle, but the set of allowed eigenvalues obtained will be the
same. Let us apply the same reasoning to our situation and let us
consider a ruler in the proper state of length $l_P$ for a given
observer, so this observer defines the rest frame for the ruler.
Now, a boosted observer will see a superposition. This means that
when measuring the length of the ruler, she might measure a
zero distance or $l_P$ or even $2l_P$ and that only the average
value of his/her measurements will be $l_P/\gamma$ and obey the
classical contraction rule. What does it mean that the observer
can see a zero distance? A priori, one would say that if two
objects/particles/excitations are distinct in space, then they
will be whatever the observer. In fact, one needs to remember that
the theory we are describing right now is an effective theory, in
a semi-classical context, with classical objects but quantum
(flat) spacetime. In a fully quantum context, one expect objects
to be described by wave packets, and is the two wave packets are
overlapping too much, then it is plausible that an observer could
not distinguish them.

In this paper we present a toy model meant to mimic closely, but
in a simplified way, the measurement of time intervals in Loop
Quantum Gravity in 3 spacetime dimensions. We then introduce boost
operators which reproduce the usual time dilatation on the
average, but such that however, as $[\hL,\hL_{boosted}]\ne 0$, an
eigenstate for an observer will become a superposition for any
boosted observer. In other words, we give an explicit realization
of the second possibility envisaged above, although in a simpler
context than the full theory, of which however we reproduce the
relevant features. Let us emphasis that information about the
length is not lost during the process of going from an observer to
the other: a pure state stays a pure state and does not become
mixed.

Our system may be thought of as a  model for a a quantum clock
(timelike rod) living
in a flat quantum geometric spacetime in which pointlike test (because
they do not affect the quantum geometry of spacetime) particles
have been introduced to give a physical meaning to the time intervals
considered. \cite{prep}

Then we map this model to a model about discrete spacelike distances
and recover the usual length contraction. Interestingly, this involves
mapping $\SU(1,1)$ to $\SU(2)$, following a proposal by 't Hooft
\cite{thooft}.

\subsection{Deformed Special Relativity and a Lorentz invariant minimal
length}

A second context in which it is interesting to try to address the
issues raised in the beginning is that of Deformed Special
Relativity theories (DSR) \cite{Giovanni, Jurek}. DSR was
especially introduced to address the second of these issues: have
a fixed invariant minimal length which can be measured by the same
experiment by any observer, and on the value of which any observer
would agree. The context is that of flat Minkowski space and this
is realized by means of a quantum deformation of the Poincar\'e
algebra of symmetries\cite{Lukierski}.

DSR can be considered \cite{GiovanniLee, ourprep} an effective flat regime
of the full quantum gravity theory, when $G\arr 0$ but $l_p$ stays
fixed, therefore $\hbar$ is sent to infinity. We are then
neglecting both the curvature of spacetime and the dynamical
aspects of the gravitational interaction, thus of the geometry,
while retaining its quantum features; it is then interpretable as
a model for the kinematics of any quantum gravity theory. It has
been indeed shown, both in the canonical framework \cite{JLL} and
in the spinfoam context \cite{3dsf}, to encode the algebra of
observables associated to a particle in $2+1$ quantum gravity,
that, we recall, is flat.

Therefore we want to consider the notion of Lorentz
contraction/dilatation also in the DSR framework, on the one hand to
see in which way the naive paradox associated to the existence of a
minimal length/time interval is solved, and on the other hand to
compare it with the LQG framework.

But what does DSR say about length contraction/dilatation?

There are many issues to understand, before being able to answer
this question in full. A first problem is in the operational
definition of what a measurement of distances is in this context.
Indeed, the speed of light is found to vary (with the energy of
the photon), so how can we measure a (spacelike) distance?
However, the most basic problem from our point of view is that it
is not clear what a length is in a DSR theory. The operators
encoding the position of a point become non-commutative, under the
assumed deformation of the Poincar\'e algebra of symmetries and of
the momentum space, obeying the commutation relations of the
$\kappa$-deformed Minkowski algebra. Therefore we simply cannot
localize a point in spacetime, so that the trivial definition of a
spacetime distance as the length of a curve connecting two points
is not available.
To tackle the problem, we see two approaches. On the one hand
we make use of the geometric picture of DSR theories based on a
curved space of momenta, and on the associated group of
symmetries. We write a (spacetime) length operator as acting on
the momentum space and study its spectrum.  Our working definition
of distance in a given direction will then be given by the very
time or space coordinate of a point on the non-commutative spacetime, and we study their spectrum and transformation properties. On the other hand, we
introduce spacetime coherent states corresponding to a point being
localized around a spacetime point $(x^0,\vec{x})$, with minimal
uncertainty. Of course, in this way the origin would assume a
privileged status, as it is in all of DSR, representing the point
of view of some observer in its rest frame. The study of the transformation properties of these states under boosts, which is left for future work, would represent another approach to the study of Lorentz contraction in DSR.
Finally, we compare the results obtained in this context with those of the
toy (loop inspired) model. It will be clear that the two approaches agree perfectly regarding the general way a
quantum discrete geometry can still be Lorentz invariant, although the details are of course different.

\section{About length spectra in Loop Quantum Gravity}

\subsection{Measuring distances and length contraction}

The most basic experiment to measure a distance between two points is to
do the ``radar''
experiment. First, one must assume that the two points/particles/objects,
$A$ and $B$,
are static with respect to one each other, so that they can effectively be
considered
as a ruler. Then the observer send a ray of light from $A$ to $B$, which
gets reflected back
to $A$. Finally, the distance $AB$ is the time $T$ of flight of the
photon:
$$
d_{AB}=\f{T}{2}c.
$$
Now the precision of the measurement is determined by two parameters: the
wavelength
of the photon $\lambda$ and the frequency $\omega$ of the clock. On the
one hand,
$\delta d \ge \lambda$ and on the other hand, $\delta d\ge c/\omega$. To
increase the
precision of the measurement, therefore, one needs to increase the energy
of both the photon and the
oscillator which we use as the clock.

Now there are strong reasons, coming from taking into account the
basic properties of gravity and the features of quantum mechanics,
to believe that one can not increase the energy localized in a
given small region of spacetime beyond the Planck scale: neither the wavelength of the photon
can be smaller that the Planck length, nor its energy can be
greater than the Planck energy, so that $\delta T \ge t_P$ and
$\delta d\ge l_P$.

From this measurement procedure, we can deduce a couple of facts.
First, spacelike lengths and space distances are not observable as
such: we effectively measure time intervals, i.e. particular
timelike lengths, so that time discreteness will imply the
discreteness of measured distances. Moreover, using a clock to
measure time intervals naturally implies that our result will
actually be discrete. And if we want to make more precise
measurements, we are stuck with the experimental resolution bound
given by the Planck time. One of the things that loop quantum
gravity achieves is to encode this bound directly in the spectrum
of time intervals, so among the very properties of the quantum
gravitational field: it is a result of the theory that time
intervals are quantized as $T=n\times t_P$ with $t_P$ as the
minimal time interval. More precisely, it is not that it is hard
to probe time intervals smaller than $t_P$: now, there doesn't
exist a time interval shorter than $t_P$.

One can refine the measurement argument by taking into account the mass of the mirrors
at the end of the ruler. Then one finds that the uncertainty (or precision) $\delta d$ actually
depends on the distance $d$ that one measures. The precise relation in 3+1d reads:
$$
\delta d \ge l_p \times \left(\f{d}{l_P}\right)^{\f{1}{3}}.
$$
If the holographic principle holds, a similar uncertainty bound is ture but with a different exponent on the right end side\footnotemark, coming from the fact that now the number of degrees of freedom scales with the area and not with the volume, as shown in \cite{viet}. In general, such refined relations read:
\be
\delta d \ge l_P \times \left(\f{d}{l_P}\right)^{\alpha},
\label{delta}
\ee
with $\alpha=1/n$ in (n+1)-d ($\alpha=1/3$ in 4d and $\alpha=1/2$ in 3d).
It should be interesting to check if we are able to recover such uncertainty bound in the
context of quantum gravity theories. As we will see in the DSR section below, the uncertainty of the measurement
of a distance actually depends on the considered distance, and we discuss how to possibly derive the previous formulae without yet being able to derive them rigourously.

\footnotetext{Roughly the argument goes as follows. Let us be in n+1 spacetime dimension and try to evaluate
the number of degrees of freedom in a volume of length $d$. Then the precision of the measurement $\delta d$ on
the distance $d$ defines our concept of point i.e of distinguishable excitations i.e of fundamental degree of freedom. The number of degrees of freedom will therefore be $(d/\delta d)^n$, but should be equal to $(d/l_P)^{(n-1)}$ if the holographic principle is realized in nature}

\medskip

Now, the usual length contraction/dilatation phenomenon is that
different observers will see different lengths $d_{AB}$ and time
intervals $T_{AB}$ depending on their speed with respect to the
ruler. More precisely, if $d_0$ is the distance seen by the static
observer (with respect to the ruler), then the length measured by
an observer with speed $v$ (in $c$ unit) is $d=d_0/\gamma$ with
$\gamma=1/\sqrt{1-v^2}\ge1$. However, as all measurements are
after all time interval measurement, it might be more interesting
to consider how these differ depending on the observer. The
relevant phenomenon is then time dilatation, i.e. the fact that an
observer moving with velocity $v$ (again in $c$ units) with
respect to one which has measured a time interval $T_0$ between
two events will measure for the same two events a time interval
$T'=T_0\gamma \ge T_0$. This also entails that relativistic clocks
will tick differently for different observers looking at them,
with different frequency that is, with the moving observer seeing
a slower clock (with frequency $1/T=1/T_0\gamma$) than the
observer at rest with respect to the clock itself (who sees a
frequency $1/T_0$).

Also in this case, however, an operational definition of space and
time intervals measurements, and a proper derivation of Lorentz
contraction and dilatation involve considering light signals sent
back and forth between two observers moving with different
velocities in spacetime and not purely geometric considerations.
In a quantum gravity context, obviously, such a careful derivation
would be even more difficult to achieve. We can therefore make us
of the fact that the purely geometric analysis of how time and
space intervals in the manifold are acted upon by the Lorentz
group of isometries of the metric gives the same results of the
physical analysis and confine ourselves to it. Of course we can
take this easier path only in the classical setting or in
simplified models of the quantum gravity regime, of which we have
no real control; hopefully, the toy model we are going to present
for time measurements and the effective framework of DSR are such
that we can safely limit the discussion to geometric
considerations only.

\subsection{A Lorentz invariant discrete time spectrum}
Let us start by writing a simple model for a quantum geometric clock in
flat space, which reproduces the relevant features of 3d loop quantum
gravity, and which gives a discrete time interval spectrum as a results of
its readings; we would like to show that this spectrum may well be
invariant under boosts, in spite of its discreteness, and that the the
usual behavior under boosts (time dilatation) is recovered when
appropriate in a quantum setting, i.e. for the mean values.

In slightly more details, we would like an operator $\wT$,
measuring time intervals in Planck time units, with eigenvectors
$|0\ra,|1\ra,|2 \ra,|3\ra\dots$, and boost operators $B_\gamma$
which can take eigenvectors to generic superpositions but such
that the average value of the boosted time interval obey (exactly)
the law of time dilatation. We work in $2+1$ dimensions to make
things simpler, but a corresponding analysis could be made for the
4-dimensional case.

Consider first, as an analogy, the expression for the spacetime
length in ordinary special relativity in a canonical basis of
coordinates, i.e. Minkowskian cartesian coordinates: $C^2 = -T^2 +
X^2 + Y^2$. What we are interested in is the equivalent of the
time interval operator $\what{T}$, because this is what changes
under boosts.

Now we have seen that in 3d Loop Quantum gravity, i.e. in a
quantum geometry in 3d flat space, the spacetime length operator
(square) is given by the Casimir of the 3d Lorentz group
$\SU(1,1)$: $C(\SU(1,1))= - J_t^2 + J_x^2 + J_y^2$, where the
$J_i$'s are the canonical generators of the Lie algebra. The
spectrum of the above is discrete or continuous depending on the
representation one works in, and, as we have said, unitary
discrete representations have then the interpretation of timelike
distances and the unitary continuous one of spacelike distances.

We are interested in the behavior of time intervals under boosts
for a given spacetime length, so let us fix a unitary
representation of $\SU(1,1)$; in this representation, of course,
boost operators and their actions are represented unitarily.

The most natural choice for the operator $\wT$, looking at the
above expression for the spacetime length operator, and keeping in
mind the analogy with the above expression for the spacetime
length in Special relativity, is the generator $J_t$ of the
rotation of $\SU(1,1)$. If one wants to have one, an intuitive
picture for this choice is to think of this operator as measuring
the spin of a spinning particle at rest, so moving along the time
axis in Minkowski spacetime, a very fundamental model for a
quantum clock indeed\footnotemark.

\footnotetext{Of course, a more carefully defined modeling of a
fundamental
quantum clock of this type should assign to it a dynamics which is
such that the system evolves monotonically through eigenstates of
increasing eigenvalue \cite{prep}. However, the dynamics of the clock and a
precise modeling of it is not relevant for the issue we want to
address in this paper.}

\medskip

Now, we choose a unitary representation of $\SU(1,1)$. A
convenient basis for our purpose is the basis diagonalizing the
operator $J_t$. More precisely, we represent $\SU(1,1)$ on the
space of $C^\infty$ functions on the circle (which play then the
role of the Hilbert space for this quantum system - the clock).
The basis is ${|n \ra}=e^{i n \theta}$, where $\theta\in[0,2\pi]$
is the coordinate on the circle and $n\in\Z$ is an integer (in
order for the functions to be single-valued). The time interval
operator $J_t$ acts as $-i
\partial/\partial\theta$ and has a discrete spectrum:
$J_t|n\ra=n|n\ra$. Then $J_\pm=J_x\pm i J_y$ will acts the left or
right translations, acting as multiplications by $e^{\pm
i\theta}$.

The exact action of the generators is:
\be
\left\{
\begin{array}{ccc}
J_t|n\ra & = & n\,|n\ra,\\
J_+|n\ra & = & \sqrt{\Om+n(n+1)}\,|n+1\ra,\\
J_-|n\ra & = & \sqrt{\Om+n(n-1)}\,|n-1\ra,
\end{array}
\right.
\ee
where $\Om$ is actually the eigenvalue of the Casimir operator $- J_t^2 + J_x^2 + J_y^2$ and
labels the different representations. The (principal) unitary
(irreducible) representations are of three kinds:
\begin{itemize}
\item the continuous series $\Om=s^2+1/4>0, s\in\R_+$: the
representation is spanned by all vectors $n\in\Z$.
\item the positive discrete series $\Om=-j(j-1), j\in\N^*$: it is
a lowest weight representation spanned by vectors $n\ge j$.
\item the negative discrete series $\Om=-j(j-1), j\in\N^*$: it is
a highest weight representation spanned by vectors $n\le -j$.
\end{itemize}

\medskip

In our toy model, we will use a $j^+$ representation. The spectrum
of the operator $\wT$ will be discrete and bounded from below: there
will be a minimal time interval of length $j$. Now let us consider
the action of boosts on this operator and on the corresponding
results of measurements. The action of boost operator $J_y$, say,
on the time interval operator is given simply by
\be
\wtl{J_t}=e^{+i\f{\eta}{2} J_y}J_te^{-i\f{\eta}{2} J_y}=
\cosh\eta J_t +\sinh\eta J_x,
\ee
while the state (of the clock) as seen by a boosted observer is
$|n \ra\,\rightarrow\,e^{i\f{\eta}{2}J_y} |n\ra$.

First the operators $J_t$ and $\wtl{J_t}$ do not commute:
\be
[J_t,\wtl{J_t}]=[J_t,\sinh\eta J_x]\ne 0,
\ee
so that an eigenvector of $J_t$ will not be an eigenvector of the
boosted time interval operator $\wtl{J_t}$: the eigenstates of the
time interval operator in one frame are not invariant under boosts
and are mapped to superpositions of eigenstates of the same
operator. In other words, a state of definite time interval size
for a given observer will be a superposition of such states for a
boosted observer, for whom the time interval will then not have a
definite value. What is {\it invariant} under boosts is the {\it
spectrum} of the operator $\wT$ measuring timelike spacetime
intervals. The continuity of the Lorentz transformation affects
the way wave function distributions transform, but not the
spectrum of the geometric quantities. We can also check that the
mean value of the time interval operator simply transforms as
expected:
\be
\la n|\wtl{J_t}|n\ra=n\cosh \eta =\cosh \eta \la n|J_t|n\ra.
\ee
As $\cosh\eta=\gamma$ for a boost, one can conclude that we
recover the usual time dilatation law on the expectation values,
and that it does not depend on the chosen $\SU(1,1)$
representation. We can go further and compute the variance of the
length state seen by the two observers. First, obviously, we have
$(\Delta J_t)^2=0$. Then:
\be
\left(\Delta
\wtl{J_t}\right)^2=\bk{n|\wtl{J_t}^2|n}-\bk{n|\wtl{J_t}|n}^2 =
\sinh^2\eta\times(\Om+n^2)=\sinh^2\eta\times(n^2-j(j-1))>0.
\ee
The fact that $\Delta\wtl{J_t}\ne0$ simply confirms that the
boosted observer sees a superposition and not an proper state of
length. Then, more interestingly, it depends on the chosen
representation and thus could be used experimentally to choose the
\lq\lq right one". Let us also point out that $\Delta\wtl{J_t}$
increases with the boost rapidity $\eta$ and with the length $n$
(so that the minimal variance is obtained for the minimal interval
$n=j$), which matches our expectations.

Notice that the boosted time interval operator $\wtl J_t$ will
have a basis of eigenstates of the same form as that given above
for the eigenstates of $J_t$, but with respect to a different
variable, and can be decomposed as above with respect to them. The
physical (or geometrical) picture is that of a cylinder centered
around the $t$-axis: the initial slice is a circle but any oblique
(or equivalently boosted) slice will be an ellipse. Now the
boosted observer will describe the state on the ellipse as a
superposition of states on her circle (located on the new boosted
slice).

This toy model realizes explicitly although in a much simplified
context the idea proposed in \cite{simone} to reconcile discrete
spectra of non invariant geometric operators in flat space loop
quantum gravity with the phenomenon of Lorentz contraction. The
analogy with the treatment of angular momentum under rotations in
ordinary quantum mechanics should be apparent.

A resulting phenomenon is that a good quantum clock, i.e one which
ticks exactly regularly, for a given observer will not be as good
for any boosted observer, or more precisely, that a boosted
observer with respect to a clock will not see it as indicating a
definite time.

\subsection{A Lorentz invariant discrete space spectrum}

We now would like to also give a Lorentz invariant toy model with
a fixed discrete length spectrum. Unfortunately, it can not be
directly inspired from 3d Lorentzian loop quantum gravity since
this theory predicts a continuous spectrum for spacelike
distances. However, what one can try to do is to somehow map the
time interval measurement to a measurement of space distances,
keeping in mind also the fact that this is after all what we
always do when measuring space distances, as discussed above.

What we are going to do then is to mimic the construction
described for time measurements in a Euclidean setting obtained by
Wick rotating the previous framework, finding an operator which
has therefore the interpretation of a distance measurement
operator, and then map the model back to the Lorentzian domain,
where the behavior of the relevant quantities under boosts is
finally read out.

The main tool we are using is a map introduced by 't Hooft
\cite{thooft} in his study of the dynamics of $2+1$ quantum
gravity, in order to study the hyperbolic geometry of spacelike
slices, formulated in terms of boost (hyperbolic) angles, using
usual angles. Another way to describe what we are doing is to say
that, by means of the 't Hooft method, we can map the $\SU(1,1)$
symmetry onto a $\SU(2)$ symmetry, and back.

We start from a Euclidean analogue of the situation described in
the previous paragraph, again inspired by Riemannian 3d Loop
Quantum Gravity. The spacetime length operator is given by the
Casimir of the $\SU(2)$ group: $C^2 = J_x^2 + J_y^2 + J_z^2$ (the
$J_i$'s are the Lie algebra generators), which has of course
always a discrete spectrum. Analogously to what we did for the
time interval operator, we choose the generator $J_z$, to
represent our toy model for non-invariant space distance operator.
One can also think of an analogous mechanical picture with a
spinning particle used to measure distances, but of course the
picture looses much physical content as we are in a Euclidean
context.

Let us consider the rotation group $\SU(2)$ and choose a (infinite
dimensional) representation of it, for example again that given by
infinitely differentiable functions on the circle. The operator
$\hL$ measuring spacelike distances will now be $J_z$ - that we
may interpret as a Wick rotated $J_t$ operator. It is of course
diagonalized easily, being represented again by a partial
derivative over the angle parameter, by the same functions on the
circle we used for $J_t$, it has discrete integer eigenvalues $n$
and corresponding eigenvectors $\mid n\ra$.

The Riemannian analogue of Lorentz boosts will now be $\SU(2)$
rotations $J_x$ and $J_y$, taken in a unitary representation of
$\SU(2)$. The simplest choice is to take a spin $j$
representation: it will contain both signs of $n$ and will have a
maximal length $j$. Eventually, we could take send $j$ to
$\infty$, or choose a infinite-dimensional representation of
$\SU(2)$.


The "boosted" space distance operator is then:
\be
\hL_{boost}=e^{-i\f{\theta}{2} J_y}\hL e^{i\f{\theta}{2} J_y}=
\cos\theta \hL +\sin\theta J_x,
\ee
where $\theta$ is the rotation parameter; again this operator
doesn't commute with the unbooosted one, so that any eigenstate of
the former is given by a linear combination of eigenstates of the
latter. Of course the "boosted" states diagonalize the "boosted"
operator, just as in the Lorentzian case. The relevant quantity
for retrieving the usual behavior under boost is the expectation
value in a given eigenstate of the space distance operator; this
expectation value transforms as
\be
\la n|\hL_{boost}|n\ra=n\cos\theta =\cos\theta \la n|\hL|n\ra.
\ee
Up to now of course we have just repeated the same analysis done
for the time interval operator, but the link with the Lorentzian
picture and with the true physical significance of boost and
Lorentz contraction is still hidden. In order to compare what we
have got so far with the situation in the Lorentzian signature we
have to map from the Riemannian to the Lorentzian setting, to Wick
rotate back. The Wick rotation map has to be of course purely
algebraic, since we are working in a coordinate free framework.
The map has to turn rotation operators $J_x$ and $J_y$ into boost
operators and relate thus the rotation angle $\theta$ to the boost
parameter $\eta$ we we were working with in the previous
paragraph.

The correct relation is:
\be
\cos\theta=1/\gamma=1/\cosh\eta,
\ee
which can also be expressed as $\tan(\theta/2)=\tanh(\eta/2)$. Let
us point out that this is a bijection between $\theta\in[0,\pi/2]$
and $\eta\in\R_+$. This is exactly the relation between angles and
boost parameters introduced in \cite{thooft} and it defines a Wick
rotation of $\SU(2)$ into $\SU(1,1)$. Written in the spinorial
representation, the map reads:
\be
e^{i\f{\eta}{2}J_y}=\cosh\f{\eta}{2}\left(1+i\tanh\f{\eta}{2}J_y\right)
\in\SU(1,1)
\quad\arr\quad
e^{i\f{\theta}{2}(iJ_y)}=\cos\f{\theta}{2}\left(1+i\tan\f{\theta}{2}(iJ_y)\right)
\in\SU(2).
\ee
In the Minkowski spacetime, we are boosting the plane $(t,x)$
sending the axis $t=0$ to the line of slope $\tanh(\eta/2)$. From
the point of view of a Riemannian spacetime, we would like then to
send the axis $t=0$ ($z=0$) to the same line of slope which
implies that the rotation angle is given by
$\tan(\theta/2)=\tanh(\eta/2)$.

Now it is easy to check that all the relations written above in
terms of $\theta$ translate into the expected relations describing
the correct behavior under boosts of the space distance operator,
of its eigenstates and eigenvectors, and of its expectation value.

In this way we introduced a length operator $\hL$ with discrete
spectrum, and boosts operators $B$ which send eigenstates $|n\ra$
of $\hL$ to superpositions $\sum_nc_n|n\ra$, but such that the
expectations values of $\hL$ and $\hL_{boost}=B^{-1}\hL B$ are
still related (exactly) by the classical Lorentz contraction law.
The variance of the boosted state will depend on the chosen
representation i.e here the value of the maximal length $j$. Let
us conclude on an interesting point: under such boosts, the pure
state $|n\ra$ will be mapped to a superposition of states $|m\ra$
where $m$ can be zero and even {\it negative}. This means we have
a kind of tunneling effect (in the sense of a classically forbidden phenomenon which is instead allowed quantum mechanically): if one observer sees a particle $A$ on
the left of the particle $B$, then there is a non-vanishing
probability that the boosted observer will see them superposed or
even the particle $A$ on the right of particle $B$.
This is the quantum counterpart of the classical phenomenon of
length contraction.

\section{Length in Deformed Special Relativity}

Deformed Special Relativity (DSR) \cite{Giovanni} was especially
introduced to address the issue of constructing a relativistic
theory, i.e. one in which inertial observers see equivalent
physics, with the equivalence being given by a group of
transformations with both a invariant speed $c$ and a invariant
length $l_P$. A feature of such a theory is now that the speed of
light depends on the energy $E$ of the beam and that $c$ is only
the speed of light as $E\arr 0$ \cite{Giovanni}. The theory is
still Lorentz invariant, even though the action of the Lorentz
transformations becomes non-linear, and the structure of the
translations is modified. The underlying symmetry algebra was
understood to be the $\kappa$-deformed Poincar\'e algebra
\cite{Lukierski}, and spacetime becomes non-commutative. Different
bases for the $\kappa$-Poincar\'e algebra give rise to different
formulations of the theory, whose physical equivalence or
difference is not clear at present. Also, and this will be a key
point in the following, a DSR theory can be characterized as a
theory of a single particle system with modified phase space, with
a curved but maximally symmetric momentum sector, and a
non-commutative flat spacetime sector. the different bases
mentioned above can be understood as different choices of
coordinates on the curved momentum sector. Many of the properties
of such a theory, including the non-commutativity of spacetime,
the curved (De Sitter) structure of the space of momenta, the
fundamental Lorentz invariance of the now quantum geometric
picture were first understood and described by Snyder
\cite{snyder}.

The general interpretation of DSR theories is that they represent
an effective description of a sector of a full quantum gravity
theory, in which one neglects the dynamical aspects of the
gravitational field, so works on a fixed quantum gravity state,
but retains at least some of its quantum properties, namely the
existence of a minimal invariant length, usually identified with
the Planck length, and also the fact that spacetime distances are
now observable properties of the gravitational field, therefore
quantum operator observables, thus in general non-commuting.
However, the precise way in which DSR theory are supposed to arise
from full quantum gravity is not clear, also since a complete
quantum gravity theory does not exist as yet. We have good
candidates for it, loop quantum gravity being one of these, and
some work on how the DSR effective description may come about from
loop gravity was done recently \cite{ourprep, JLL, 3dsf}, with the most
solid results having been obtained in the 3d case \cite{JLL,3dsf}.

Let us note that DSR is not yet fully understood and that there
exists different proposals about the physics of the theory. In
this context, one should view our analysis as an attempt to
clarify a few points of the corresponding physical theory, and
more precisely about the quantum geometric structure of spacetime
underlying it. We discuss first measurements of distances in the
DSR framework, and the problems that an operational definition
presents. Second we give a working geometric definition of
distances and distance quantum operators, study their spectral
properties and their behavior under boosts, investigating what
happens to length contraction in this context. Finally we analyze
the algebraic structure of the mathematical theory in order to
understand how to localize points in a non-commutative spacetime.
For this purpose, we introduce wave packet states to represent the
best possible definition of points of a quantum spacetime, and
study some of their properties. All this will be done in the two
most common bases (or versions) of DSR theory: the Snyder basis
and the kappa-Minkowski basis.

\subsection{Distance measurements in VSL theories}

Let us consider distance measurements in a DSR theory, using the
same radar experiment as usual. DSR theories, in some choices of bases, predict a varying speed of light (VSL): the speed of light depends
on the energy of the photon (it converges to $c$ only at low energy $E\arr 0$ and goes to $\infty$
when $E$ reaches the Planck energy). When this does not happen, the usual operational definition as in special relativity applies, and the difficult issues are those resulting from the non-commutativity of spacetime coordinates, i.e. how to define spacetime distance operators and spacetime points. When this happens, instead, there are clearly additional complications, since the speed of light $c$
is use to define the unit of length.

In more details, simply measuring the time of travel of the photon
will give us an effective distance:
\be
d_{eff}(E)=\f{T_{(E)}}{2}\times c,
\ee
which would be different from the ``true'' distance:
\be
d=\f{T_{(E)}}{2}\times c(E).
\ee
Then assuming that, generically, the speed $c(E)$ increases with
the energy $E$ (with $c(E\arr 0)=c$ and $c(E\arr E_P)=\infty$),
$d_{eff}$ would decrease even though $d$ should remain constant
(as a property of the physical system and not of the measurement).

As result, the measurement of a single distance does not make sense. Only
ratios of distances contain valuable information. Working at a
given energy $E$, one can compare the length of two different
rulers, since their ratio will not depend on the speed $c(E)$.
Of course, apart from the more involved context, this statement is not
valid only for  DSR but also
applies to ordinary special relativity.

Now, one might want to get the curve $c(E)$. There is no real way
to get $c=c(E\arr 0)$, and we can once again only have access to
the ratio $c(E')/c(E)$. At this point, we should also remember
that we have a natural uncertainty in the distance measurements:
$$
\delta d_{eff}(E) \ge \lambda(E),
$$
with $\lambda(E)$ goes to $\infty$ when $E\arr 0$ and decreases
when $E$ grows. One does a more precise distance measurement by
using a beam of light of higher energy. As a result, we naively
face a problem: how to get a precise measurement of the speed of
light $c(E\arr 0)$? We just have to use a different reference
energy $E_0$ such that $\lambda(E_0)$ is reasonable, and then we
only compute distance or speed ratios, so that it doesn't really
matter which reference energy we use.

\medskip

Now, let us have a look at the Lorentz length contraction and
discuss the measurement of the length of a ruler made by
different observers. Now DSR theories are understood to still be
Lorentz invariant. However, the representation of the Lorentz
transformations is modified (and becomes non-linear). This means
that, although we do not expect a modification of the physical
phenomenon, we still do expect a modification of the relation
between the speed $v$ of the boosted observer and the contraction
factor $\gamma$ (or equivalently between the speed $v$ and the
boost rapidity or angle $\eta$). Moreover, this is assuming that
the two observers agree on the energy they will use. Else they
will use different speeds of light and thus obtain different
(effective) distances. For example, in the special case where the
two observers use the same beam of light, the energy of the light
beam will not be the same in the two reference frames and that
will lead to an extra correction to the Lorentz contraction law.

Thereofore, we have here two
possible procedures to test the predictions of DSR. Both depend on the
details of the chosen DSR theory. The first is the measurement of
the curve $c(E)/c(E_0)$. And the second is the law of length
contraction. Looking only at the first experiment will only
calibrate the function $c(E)$ and tell us we are in a VSL theory.
Only a consistency check between the results of the two
experiments could be considered as a first check of the DSR
framework.

We shall let aside these operational considerations from now on and tackle the issue of distance measurements in a more abstract and purely quantum geometric way.

\subsection{A Lorentz invariant DeSitter lattice structure for spacetime}

The first version of what we now call Deformed (or Doubly) special
relativity was introduced by Snyder as an example of a Lorentz
invariant {\it and} discrete quantum spacetime
\cite{snyder}. The idea is to consider the spacetime coordinates
as operators, with discrete spectra, acting on a curved space of
momenta, where this curvature governs the operator nature and
therefore the non-commutativity of spacetime. More precisely,
momentum space is described as a smooth manifold with De Sitter
geometry, also understood as the homogeneous space
$\SO(4,1)/\SO(3,1)$, therefore with a transitive action of the
$\SO(4,1)$ group; this symmetry group is then decomposed according
to the Cartan decomposition into the Lorentz subgroup $\SO(3,1)$,
with canonical action on the homogeneous space, and the remainder,
generated by operators $x_1,x_2,x_3,t$ giving translation on the
De Sitter space, and interpreted as coordinates of a dual flat
4-dimensional spacetime, which then results in being
non-commutative\footnotemark.

\footnotetext{We focus here on the 4-dimensional case, because it is the one of greatest physical interest, for reasons that should be obvious. DSR theories can of course be easily generalised to any dimension having as momentum space the higher De Sitter constructed as homogeneous space $SO(n,1)/SO(n-1,1)$. Keeping this in mind, it is easy to realize that our toy model presented in the previous section can be seen formally as a kind of (Riemannian) 2d DSR.}

If one wants a coordinate presentation of the above, the dS space of
momenta is defined as embedded in a 5-dimensional Minkowski space by
$$
dS=\{\eta^2_0-\eta^2_1-\eta^2_2-\eta^2_3-\eta^2_4=-\eta^2\},
$$
and the spacetime coordinates are defined as:
$$
x_i=ia\left(\eta_4\f{\pp}{\pp\eta_i}-\eta_i\f{\pp}{\pp\eta_4}\right)
\qquad
t=x_0=ia\left(\eta_4\f{\pp}{\pp\eta_0}+\eta_0\f{\pp}{\pp\eta_4}\right),
$$
where $a$ is a length. $a$ is usually taken as the Planck length
$l_P$, but can be more simply considered as the length resolution
of a particular observer \cite{ourprep}. The Lorentz algebra is
undeformed and the generators are explicitely given by
$$
J_{ij}=i\hbar\left(\eta_j\f{\pp}{\pp\eta_i}-\eta_i\f{\pp}{\pp\eta_j}\right),
\qquad
K_i=i\hbar\left(\eta_0\f{\pp}{\pp\eta_i}+\eta_i\f{\pp}{\pp\eta_0}\right),
$$
where $J_{ij}$ and $K_i$ are the rotation and boost generators of
the $\SO(3,1)$ subgroup, respectively. They act as usual on the
spacetime coordinates (the action is NOT deformed), but the {\bf
spectrum of the $x_i$ operators is discrete} and is $\Z a$, while
the spectrum of $t$ is still $\R$. Thus {\bf $a$ is an universal
length scale seen by all observers}. Nevertheless, as a
consequence of these definitions, spacetime becomes
non-commutative and the coordinate commutators are:
\be
[x_i,x_j]=\left(i\f{a^2}{\hbar}\right)J_{ij},\qquad
[t,x_i]=\left(i\f{a^2}{\hbar}\right)K_i.
\ee

All operators act on the homogeneous space $\SO(4,1)/\SO(3,1)$, so
that the Hilbert space of our DSR theory is the space of $L^2$
functions on $\SO(4,1)/\SO(3,1)$. It is then natural to introduce
the conjugate momenta operators $p_i,p_t$ as a choice of
coordinates on the dS space\footnotemark.
\footnotetext{Mathematically, the choice of momenta is arbitrary,
it will need to be physically motivated. Nevertheless, a canonical
choice is:
$$
p_i=\f{\hbar}{a}\f{\eta_i}{\eta_4}, \qquad
p_t=\f{\hbar}{a}\f{\eta_0}{\eta_4}.
$$
Then the commutation relations are
$$
[x_i,p_i]=i\hbar\left(1+\left(\f{a}{\hbar}\right)^2p_i^2\right),
\quad
[x_i,p_j]=i\hbar\left(\f{a}{\hbar}\right)^2p_ip_j,
\quad
[t,p_t]=i\hbar\left(1-\left(\f{a}{\hbar}\right)^2p_t^2\right),\dots
$$
Another very interesting choice, in a sense closer in spirit to
the point of view advocated here, is presented and discussed in
\cite{JurekDeSitter}, and it is basically given by the usual
orispherical coordinates on the De Sitter hyperboloid, following
from the Cartan decomposition of the $\SO(4,1)$ group \cite{VK}.}
They act multiplicatively, and have a continuous spectrum.  The
commutators between the $x$'s and the $p$'s get corrections in
$a^2$, so that the action of the Poincar\'e translations is
deformed. Nevertheless, when $|p|\ll\hbar/a$, then we recover the
usual quantum phase space\footnotemark.


\footnotetext{One can easily compare the present situation with the
one in a straightforward effective quantization of the flat space
of ordinary special relativity, where one has a flat space of
momenta on which the Poincar\'e group of symmetries $ISO(3,1)$
acts transitively, and the coordinates of spacetime can be
identified with the generators of the translation part of this
group; the resulting quantum flat spacetime is commutative, since
the algebra of translations is abelian, and the space distance and
time interval operators have all continuous spectrum.}

In this context of a non-commutative spacetime with discrete
coordinates, we want to discuss the behavior of measured distances
and time intervals under Lorentz boosts, and the phenomenon of
length contraction. Of course, in order to do this, we have to
give a definition of spacetime lengths, space distances and time
intervals. As we have discussed, an operational definition,
already quite difficult to give in a general quantum geometric
context, is made even more tricky here in a DSR context by the
fact that the velocity of light is not constant. However, we can
rely on the algebraic and geometric picture outlined above to give
a reasonable, albeit formal, definition: the very spacetime
coordinate operators are interpreted as (non-invariant) time or
space distances, describing possible geometric measurements
carried out by an observer living on a non-commutative spacetime,
to whom this choice of coordinates is relative.

Accordingly, the spacetime length operator (square), or invariant distance
(square), is given by the operator:
\be
\what{L}^2\,=\,-\, t^2\,+\,x_1^2\,+\,x_2^2\,+\,x_3^2
\ee
which is simply the difference of the two quadratic Casimirs of
$\SO(4,1)$ and $\SO(3,1)$, as appropriate for an invariant operator
defined on the coset $\SO(4,1)/\SO(3,1)$, and has the familiar form
of a flat space spacetime interval. Note also that from the
geometric point of view this operator is simply the d'Alambertian
invariant operator on De Sitter space.

Let us start by considering its spectrum. As we work in the
$p$-polarisation and we consider the $x$'s are (translation)
operators on the hyperboloid, the wavefunctions are functions on
De Sitter space, i.e. the Hilbert space of the theory is given by
$L^2$ functions on De Sitter space. A basis of such functions is provided by the
canonical basis in the vector space of simple representations of
$\SO(4,1)$, i.e those who have a $\SO(3,1)$-invariant vector (see
in appendix for details about the representation theory). These
representation are labeled by either an integer parameter $n$ or
by a real positive parameter $\rho$. More precisely, let us choose
a simple representation ${\cal I}$ of $\SO(4,1)$ and note $v$ a
$\SO(3,1)$ invariant vector and $w$ another arbitrary vector in
the same representation. Then the function:
\be
f_{v,w}^{({\cal I})}(g\in\SO(4,1))=\la w|D^{{\cal I}}(g)|v\ra
\ee
is truly a function on the coset $\SO(4,1)/\SO(3,1)$, thus on De
Sitter space. Such functions generate, varying $w$, the space of
$L^2$ functions over De Sitter space, as any function on
$\SO(4,1)/\SO(3,1)$ can be decomposed into a linear infinite (as
the representation of $\SO(4,1)$ in the space of functions on the
above coset is infinite dimensional) combination of such basis
functions by harmonic analysis (see \cite{VK, Gelfand, Ruhl}), this
decomposition involving a sum over the discrete parameter $n$ and
an integral over the real parameter $\rho$.

Now, choosing a vector $w_{{\cal L}}$ living in a given $\SO(3,1)$
representation ${\cal L}$ within ${\cal I}$, it is easy to check that
the length operator will act diagonally on  $f_{v,w}$:
\be
\what{L}^2 |f_{v,w_{{\cal L}}}^{({\cal I})}\ra=
\left[C_{\SO(4,1)}({\cal I})-C_{\SO(3,1)}({\cal L})\right]
|f_{v,w_{\cal L}}^{({\cal I})}\ra.
\ee


For a $\SO(4,1)$-representation ${\cal I}=(0,\rho)$ from the
continuous series, it decomposes onto $\SO(3,1)$-representations
of the type $(0,\tau\in\R_+)$, more precisely (see \cite{LN} for the full analysis),
$$R^{(0,\rho)}_{\SO(4,1)}=2\int_0^\infty
d\tau\,R^{(0,\tau)}_{\SO(3,1)},$$ and the
corresponding (spacetime) length eigenvalues are:
\be
\what{L}^2\, | f_{v,w_{\tau}}^{\rho}\ra
\,=\,
a^2\left[-\rho^2-\f{9}{4}+\tau^2+1\right]\, |
f_{v,w_{\tau}}^{\rho}\ra.
\ee
A representation ${\cal I}=(n,0)$ from the discrete series get
decomposed \cite{LN} as:
\be
R^{(n,0)}_{\SO(4,1)}=
\bigoplus_{m=0}^{n}R^{(m,0)}_{\SO(3,1)}+
\int_0^\infty d\tau\,R^{(0,\tau)}_{\SO(3,1)},
\ee
so that the corresponding (spacetime) length eigenvalues are\footnotemark:
\bes
\what{L}^2\, | f_{m}^{n}\ra
&\,=\,&
a^2\left[n(n+3)-m(m+2)\right]\, | f_{m}^{n}\ra, \\
\what{L}^2\, | f_{\tau}^{n}\ra
&\,=\,& a^2\left[n(n+3)+\tau^2+1\right]\, | f_{\tau}^{n}\ra.
\ees
Notice that due to the constraint $m\le n$, $L^2({\cal I}=n)$ is
always positive, and thus the discrete simple representations
correspond to time-like spacetime intervals.


\footnotetext{This is for to the 4-dimensional case; in the general d-dimensional case the invariant spacetime length operator is given by $\what{L}^2= C_{\SO(d,1)}({\cal I}) - C_{\SO(d-1,1)}({\cal L})$, the representations of the relevant groups are again labeled by either an integer or a real parameter, the decomposition into irreps of the subgroup is analogous and the corresponding eigenvalues are:  $a^2\left[ -\rho^2 - \left(\frac{d+1}{2}\right)^2 + \tau^2 + \frac{d^2}{4}\right]$ in the case of continuous representations of $SO(d,1)$, and $a^2\left[ n( n + d - 1 ) - m ( m + d -2 ) \right]$ or $a^2 \left[ n ( n + d - 1) + \tau^2 + \frac{d^2}{4} \right]$ for the discrete series.}

Notice that, overall, there is no real discreteness of spacelike
or timelike spacetime intervals. Moreover, as a same value of
$L^2$ can be obtained using different states, one can wonder about
the physical (or geometrical) interpretations of these states.
Indeed maybe we only use a particular subspace of states when
dealing with an experiment measuring the length of a ruler. Or
more generally, it is likely that the particular physical
phenomenon under consideration will dictate us which class of states we need to use
(to get more physical insight, it would be interesting to compute the average
values of the $x$ spacetime coordinates).
Then we will get different behaviors of the length operator. For
example, if we restrict ourselves to a fixed $\SO(4,1)$
representation $n$, then we are only dealing with positive $L^2$
i.e spacelike intervals, and moreover it turns out that distances
smaller than $n\times a$ are quantized while distances larger than
$n\times a$ remain continuous. A potential use of this choice of
subspace is when one wants to fix a semi-classical length scale
above which spacetime looks classical and smooth but beyond which
the quantum structure of spacetime becomes relevant. The other
example is when one selects the subspace defined by a fixed
$\SO(4,1)$ representation $\rho$, the length spectrum is always
continuous but there now exists a maximal timelike interval of
length $\rho\times a$: it could be useful when studying
(geometrical) properties of a (gravity) system in a finite range
of proper time.

\medskip

One can also look the action of other operators such as the
coordinates. These, as we said, have the interpretation of
non-invariant (abstract) space distance measurements, giving the
possible locations of an object with respect to the observer to
which the Snyder's choice of coordinates is associated. To
diagonalise the operator $x_\mu$, one uses the basis diagonalising
$J_{4\mu}$ in the representation ${\cal I}$.
As it is impossible to diagonalise the generators $J_{4\mu}$ simultaneously,
one can never diagonalise the spacetime coordinates
simultaneously, as it is also clear from the commutation relations
among them. This of course does not prevent
from defining space and time distances separately, although their
operational meaning is tricky (to say the least) due to this
non-commutativity. We have already mentioned this definition above
and given the spectrum of the coordinate operators, defining
possible results of these hypothetical measurements. However, it
is clear that the $x_i$'s, as we have mentioned above, have
discrete (equispaced) spectra, while the $t$ operator, being generator
of a boost, has a
continuous one, whatever irreducible representation one chooses
for $SO(4,1)$ acting on the De Sitter space of momenta.

One can also introduce the notion of a {\it space distance}
$\what{l}^2=x_ix_i$, although this cannot have the meaning of a
space distance at a given time, since measuring the former to a
given precision requires a large indeterminacy in the latter: as
$[t,x_i]\ne0$, knowing {\it where} an event occurred implies not
knowing {\it when} it did!. This space distance operator
$\what{l}$ is a Casimir of the $\SO(3)$ subalgebra of
(space) rotations and has a discrete spectrum. In order to
diagonalise $\what{l}$, one needs to decompose the $\SO(4,1)$
representations ${\cal I}$ into $\SO(3)$ representations\footnotemark.
 Let us underline the somehow counter-intuitive fact that
$[x_i,\what{l}^2]\ne 0$.

\footnotetext{A simple $\SO(4,1)$ representation can be decomposed into simple
$\SO(3,1)$ representations. Then a $\SO(3,1)$ representation $(n,\rho)$ decomposes as:
$$
R^{(n,\rho)}_{\SO(3,1)}\,=\,\bigoplus_{j=n}^{\infty} R^j_{\SO(3)}.
$$}

\medskip

Now, having set up the basic quantum geometric setting for space
an time measurements, one can start investigating the behavior of
these operators under Lorentz boosts, and how the usual phenomena
of length contraction and time dilatation look like. It will be
clear that the situation is very much analogue to the one we
pictured in our toy model of time measurements, inspired by Loop
Quantum Gravity. We concentrate on space measurements and length
contraction, since it is the discreteness of geometric spectra
that we are most concerned with, when dealing with Lorentz
invariance. The analogous analysis can be performed for the time
measurements as well.

To discuss the issue of Lorentz contraction, we needs to compare
operators with their boosted counterpart. To start with, let us
imagine that an observer has a ruler in the direction $x_1$, with
one end of the ruler in the same position as the observer and of
course at rest with it. The
states of the ruler can be decomposed on eigenstates of
$\what{x_1}$. On the other hand, a boosted observer will decompose
the ruler states on eigenstates of $x_1^{(boosted)}\equiv e^{i\eta
K_1}x_1e^{-i\eta K_1}$. Then it is straightforward to check that:
\be
[x_1^{(boosted)},x_1]=[\cosh\eta \, x_1+\sinh\eta \,x_0,x_1]=
\sinh\eta\left(\f{a^2}{\hbar}\right)K_1\ne 0.
\ee
This is the same feature as introduced in our toy model: a proper
state of length for a given observer can not be a proper state for
a boosted one. Notice in fact that the toy model
can be seen formally as a kind of 2-dimensional version of DSR, so it
is not so surprising that the behavior of geometric operators under
boosts is similar. Now, telling which state will the boosted observer
see given the initial state is a more involved problem with respect to the to toy model presented in the previous section, due to the more complicated action of the other coordinate operators on the eigenstates of one of them. Therefore it is not as straightforward as it was in that case to check that we recover the usual length contraction (or time dilatation) law for the mean values of the space (or time) distance operators, and a more detailed analysis would be needed. However the general picture is the same and it is again the non-commutativity shown above that allows to maintain Lorentz invariance in this quantum setting in the presence of discreteness of geometry.

The first
issue to solve to have a complete treatment of such problem would be to define time slices, in order to define the spaces
associated to each observer and write the state of the ruler
projected onto the respective slices. Of course, as time doesn't
commute with the space coordinates, we will need a concept of
approximate time slices. More generally, one would also like
to define approximate points: a class of coherent states,
representing semi-classical spacetime points, minimizing the
coordinates uncertainty relations; how this can be done will be
sketched later in this section.

Defining boosted time slices requires computing the action of
boosts over the coordinate $x_0$:
$$
x_0^{(boosted)}=\cosh\eta x_0 +\sinh\eta x_1,
\qquad [x_0^{(boosted)},x_0]\ne0.
$$
A ideal time slice would be defined as an eigenspace of $x_0$
corresponding to a given eigenvalue.
Now, it is clear that we need to thicken these slices in order
to be able to localize space coordinates, due to the commutation relations between them. Let us call such
subspaces $T(t), t\in\R$. We also define the time slices
$T^{(b)}(t)$ relative to the boosted observer and its time
coordinate $x_0^{(boosted)}$. A natural requirement (and easy to
satisfy) is that boosts send $T(t)$ onto $T^{(b)}(t)$: here
$T^{(b)}(t)=e^{i\eta K_1} T(t)$. Then let us pick up a state
$|\Psi\ra$. The state as seen by the first observer will be the
projection of $|\Psi\ra$ onto $T(t=0)$ and we will be looking at
the average of the operator $x_1$. The boosted observer will see
the projection of $|\Psi\ra$ onto $T^{(b)}(t=0)$ and will consider
the average of $x^{(b)}_1$. To reproduce the exact setting of the
Lorentz length contraction, we would need to choose a state picked
around a value of $x_1$ and completely delocalized in $x_0$ i.e representing a
particle trajectory (the worldline of the end of the ruler).


To make the argument more generic, one can work with the
introduced space length operator $\what{l}^2$ in which case it is
straightforward to check that
$[\what{l}^2_{(boosted)},\what{l}^2]\ne0$.

\medskip

Let us now come to another subtle point concerning the quantum
geometry of this non-commutative spacetime setting, related to the previous one of constructing time slices. What is a
spacetime point? How can an observer localize any event, even in
principle, let alone the operational specification of such a
localization procedure?

Clearly, due to the commutation relations among spacetime coordinates,
that we identified with space and time distance measurements with
respect to a point (origin) where the observer using such a coordinate
system is, it is not possible to assign exact values for all these
coordinates; this means that no system can be described by a state
which is simultaneously eigenstate of all of these operators, no
system can be perfectly localized in space and time, although its
spacetime distance from the observer can be given exactly (the state
describing this system may be labeled by a given irreducible
representation of the $\SO(4,1)$ group).
What is then the state of a
system, if we have localized it as much as the framework allows us to
do? This is a familiar problem in Quantum Mechanics, where classically
commuting observables do not commute at the quantum level (think for
simplicity of coordinate and momentum in a given direction), and it
can be dealt with in the usual way: such a state must be given by a wave
packet in spacetime, for a given
spacetime distance, i.e. by a state living in a given representation
of $\SO(4,1)$, and that is peaked around some eigenvalue of the space
and time coordinate operators with a given small dispersion.
This would be our definition of a \lq\lq spacetime point" in this non-commutative setting.

Let us now discuss the construction of these states.
The relevant commutation relation is $[ x_\mu , x_\nu ] = i\f{a^2}{\hbar} J_{\mu\nu}$.
Focusing on a particular space coordinate $x_i$ and the time coordinate $t$,
the relative indeterminacy in space and time coordinates follows:
    \be
    \delta_\Psi t\,\delta_\Psi x_i\,\geq\,-\f{i}{2}\la [ t, x_i ]
\ra_\Psi\,=\,\f{a^2}{2 \hbar}\la K_i \ra_\Psi.
    \ee
This is the only equation limiting the values we can assign
simultaneously to the coordinates $x_0$ and $x_i$ in the given
state $\mid \Psi \ra$. We see that there is a fundamental
indeterminacy as soon as the state is such that the boost operator
$K_i$ has a non-zero mean value in it. One could think of fixing this mean value
$\la K_i \ra_\Psi$, say equal to $k$, and treat it as a fixed datum in determining the dependence of the state $\Psi$ itself on the coordinate operators $t$ and $x_i$, building it in such a way as to minimize the above uncertainty relation with a constant $\frac{a^2 k}{2\hbar}$ on the right end side. Then we could write coherent states as for the harmonic oscillator. Unfortunately, it doesn't seem that it provides a good approximation for solving the problem and we will need to take into account the dependence of $\la K_i \ra_\Psi$ on the state $\Psi$.
This easier shortcut being unavailable, we should then use the tools of coherent states for Lie groups (see for example \cite{cohstates})
to write down coherent states for localized spacetime points, defining them in this case as those coherent states for the group $SO(4,1)$ invariant under the $SO(3,1)$ subgroup that are closest to the classical values. We give the general ideas of this construction and some examples of it in the following subsection.


\subsection{Lie groups' coherent states to localize points}
A general construction for all Lie groups can be found in \cite{cohstates}.
The procedure for constructing a system of coherent states is as follows: given a Lie Group $G$ and a representation $T$ of it acting on the Hilbert space $\cal{H}$, with a fixed vector $\mid \Psi_0\rangle$ in it, the system of states $\left\{ \mid \Psi_g\rangle = T(g)\mid \Psi_0\rangle\right\}$ is a system of coherent states.
The point is then to identify which are the states  $\Psi_0$ that are closest to classical, in the sense that they minimize the invariant dispersion or variance $\Delta = \Delta C = \langle \Psi_0 \mid C \mid \Psi_0 \rangle - g^{ij} \langle \Psi_0 \mid X_i \mid \Psi_0 \rangle \langle \Psi_0 \mid X_j \mid \Psi_0 \rangle$, where $C=g^{ij}X_i X_j$ is the invariant quadratic Casimir of the group $G$ (that, remember, we have taken to represent the invariant spacetime length in our quantum geometric setting).
The general requirement is the following: call $\lalg{g}$ the algebra of $G$, $\lalg{g}^c$ its complexification, $\lalg{b}$ the isotropy subalgebra of the state $\mid \Psi_0\rangle$, i.e. the set of elements $b$ in $\lalg{g}^c$ such that $T_b \mid \Psi_0\rangle = \lambda_b \mid \Psi_0\rangle$, with $\lambda_b$ a complex number, and $\bar{\lalg{b}}$ the subalgebra of $\lalg{g}^c$ conjugate to $\lalg{b}$; then the state $\mid \Psi_0\rangle$ is closest to the classical states if it is most symmetrical, that  is if $\lalg{b}\oplus\bar{\lalg{b}} = \lalg{g}^c$, i.e. if the isotropy subalgebra $\lalg{b}$ is maximal.
This construction is covariant and indeed completely general, and it applies also to the case of our present interest, i.e. constructing coherent states for the group $SO(4,1)$ invariant under the subgroup $SO(3,1)$. However, to construct these states explicitly is no trivial task.

Let us start by giving the example of $\SU(2)$ coherent states. The goal
is to minimize the uncertainty relations $\delta J_i.\delta J_j \ge \la J_k \ra$
or equivalently the invariant variance $\Delta=\la J_iJ_i\ra-\la J_i\ra\la J_i\ra$.
 For $\SU(2)$,
the coherent states (or semi-classical states) are the states of highest weight i.e
in the standard basis $|j,m=\pm j\ra$ (for any choice of $z$-axis) and the rotated
$g|j,m= j\ra$ for all $g\in\SU(2)$. These same states can also be constructed out of the representation of $\SU(2)$ as harmonic oscillators\footnotemark.
The fuzzyness of the state is then quantified by:
\be
\Delta=\la J_iJ_i\ra-\la J_i\ra\la J_i\ra=j(j+1)-j^2=j.
\ee
Computing the different components for $|j,j\ra$, we get $\Delta_z=\la J_z^2\ra-\la J_z\ra^2=0$
and $\Delta_x=\Delta_y=j/2$.
And it is straightforward to check that they minimize the Heisenberg-type uncertainty relations.
More precisely, these states $g|j,m= j\ra$ can be seen as semi-classical states on the 2-sphere ${\cal S}_2$
of radius $j$. Interpreting the distance as given by  $l=j\times l_P$ and the uncertainty of the measurement as given by $\delta l=\sqrt{\Delta}l_P$, we are finally lead to:
\be
\delta l=\sqrt{j}l_P=l_P\sqrt{\f{l}{l_P}}.
\ee
Hence we have recovered a similar relation of the expected shape \Ref{delta} with $\alpha=1/2$, which
is supposed to correspond to the 3d case.

\footnotetext{$\SU(2)$ can be represented in terms of two harmonic oscillators $a,a^\dagger$ and $b,b^\dagger$:
$$
J_z=\f{1}{2}(a\dag a-b\dag b), \qquad J_+=a\dag b, \qquad J_-=J_+\dag=ab\dag.
$$
The spin $j$ is given by the Casimir operator $N=\f{1}{2}(a\dag a+b\dag b)$ which commutes which the $J$'s.
Coherent states for the system of two harmonic oscillators are:
$$
|z_az_b\ra=\sum_{n_a,n_b}\f{z_a^{n_a}}{n_a!}\f{z_b^{n_b}}{n_b!}|n_an_b\ra,
$$
and they can projected onto a spin $j$ representation by imposing that $N=(n_a+n_b)/2=j$. The resulting state can be parametrized solely by $\xi=z_a/z_b\in\C$ and the mean values of the $J$'s are:
$$
\la J_z\ra=j\f{1-|\xi|^2}{1+|\xi|^2},\quad
\la J_x\ra=j\f{2Re(\xi)}{1+|\xi|^2},\quad
\la J_z\ra=j\f{2Im(\xi)}{1+|\xi|^2},
$$
which is simply the parametrization of the 2-sphere of radius $j$ as the Riemann sphere.}

Interestingly, we recover the same relation in the Lorentzian case using $\SU(1,1)$ instead of $\SU(2)$.
The invariant variance is given by $\Delta=\la J_t^2-J_x^2-J_y^2\ra-\la J_t\ra^2+\la J_x\ra^2+\la J_y\ra^2$.
Some semi-classical states minimizing $\Delta$ are given by the lowest weight states of the discrete series of representations. Such representations are labelled by an (half-)integer $n\ge 1$ (and a parity $\pm$), and the states
are the $|n,m=n\ra$ (diagonalizing $J_t$ with eigenvalue $m=n$) for all choice of $t$-axis. These states can be interpreted as semi-classical states on the upper part of the 2-sheet hyperboloid in the 3d Minkowski space of radius $n$. Then
$$
\Delta=n(n-1)-n^2=-n,
$$
which gives an uncertainty $\delta l=\sqrt{|\Delta|}l_P=\sqrt{n}l_P$ on the distance $l=n\times l_P$.




Then we should use the same techniques of Lie group coherent states to build coherent states localizing points in 3d and 4d DSR. Here the explicit identification of the coherent states closest to classical is more complicated, and the non-compactness of the isotropy subgroup makes the details of the construction even more involved. The goal would be to minimize the fuzzyness
$\Delta=\la x_\mu x_\mu\ra-\la x_\mu\ra\la x_\mu\ra=\la \what{L}^2\ra -\la x\ra^2$. The first step would be to compute the mean values of the $x_\mu$ on the spacetime length eigenvectors. After a careful analysis, we would then be able to extract the uncertainties $\delta l$ in the measure of distances in DSR and check whether they have the expected behavior. For example, taking states $|f_m^n\ra$, {\it assuming} that the expectation values of the $x_\mu$ vanishes, i.e that they are states centered around the origin, then semi-classical states would be defined as minimizing the spacetime length: it would be states with $m=n$. Then $\delta x=\sqrt{\Delta}=l_P\sqrt{n}$ which has the same behavior as the $\SU(2)$ coherent states. At the end of the day, we expect that such a notion of semi-classical coherent states for DSR will always give $\delta l=l_P\sqrt{l/l_P}$ with $\alpha=1/2$ and never an exponent $\alpha=1/3$\dots but this is only a conjecture. Further analysis is needed to confirm this and to construct the coherent states exactly\footnotemark.

\footnotetext{For this purpose, we think that the representation of the algebra $so(5)$ in terms of harmonic oscillators should be useful to write coherent states, the generators being expressed using two sets of creation-annihilation operators $a_{1,2},a^\dagger_{1,2}$:
$$
X_{ij}=a_ia_j,\quad X^{ij}=a^\dagger_ia^\dagger_j,\quad
X_l^k=\f{1}{2}\left(a^\dagger_ka_l+a_la^\dagger_k\right).
$$ }

Let us stress the full covariance of this, albeit complicated, construction of coherent states minimizing the uncertainty relations.

\subsection{About an AntiDeSitter lattice structure}

Snyder's idea can be also applied to the homogeneous space defined
as $\SO(3,2)/\SO(3,1)$ i.e the AntiDeSitter space. Once again the
spacetime manifold is recovered as the tangent space to the
hyperboloid. However, due to the change of signature, the
operators $x_i$ corresponding to the space coordinates become
non-compact (anti-hermitian) while $t$ is represented by a compact
generator (hermitian). Therefore, {\bf space coordinates have a
continuous spectrum while time gets quantized}. Let us point out
that this statement is similar to results from the spin foam
approach to quantum gravity \cite{Daniele}.

The Hilbert space of the theory will be the space of $L^2$
functions over the AdS space, which can be generated as previously
using simple representations of $\SO(3,2)$ (i.e containing a
vector invariant under the $\SO(3,1)$ Lorentz subgroup). Now the
spectrum of spacetime lengths will be given by the difference of
the $\SO(3,2)$ Casimir and the $\SO(3,1)$ Casimir. Nevertheless,
as the signature changed, the sign of the $\SO(3,1)$ Casimir is
also changed.


Simple representations of $\SO(3,2)$ are once again of two types,
either labeled by an integer $n\in\N$ or by a real parameter
$\rho\in\R_+$ (and a parity $\epsilon=\pm$). But their
decomposition into $\SO(3,1)$ representations is different than in
the deSitter case\cite{LN}:
\bes
R^n&=&\int_0^\infty d\tau\, R^\tau, \\
R^{\rho,\pm}&=&2\int_0^\infty d\tau\,R^\tau\oplus
\bigoplus_{m=0}^\infty R^m,
\ees
so that the spacetime length eigenvalues are\footnotemark:
\be
\f{L^2}{a^2}\,=\,
-\left[n(n+3)+\tau^2+1\right]
\quad\textrm{or}\quad
\rho^2+\f{9}{4}-\tau^2-1
\quad\textrm{or}\quad
\rho^2+\f{9}{4}+m(m+2).
\ee

\footnotetext{Again, this is for the 4-dimensional case; in the d-dimensional one, the analogous decomposition gives the eigenvalues: $- a^2\,\left[ n ( n + d - 1) + \tau^2 + \left( \frac{d- 2}{2}\right)^2\right]$ or $a^2\left[ \rho^2 + \left( \frac{d- 1}{2}\right)^2 - \tau^2 -  \left( \frac{d- 2}{2}\right)^2\right]$ or $a^2\left[ \rho^2 + \left( \frac{d- 1}{2}\right)^2 + m ( m + d - 2)\right]$, depending on the representation of $SO(3,2)$ we are decomposing.}

The same discussion as for the deSitter case applies here. Indeed
we notice that, overall, there is no real discreteness of
spacelike or timelike spacetime intervals. More precisely, as a
same value of $L^2$ can be obtained using different states, one
can wonder about the physical (or geometrical) interpretations of
these states, and we expect than different physical phenomenon
will be analyzed using different subspaces of the space of states. A interesting
case is when restricting to a fixed $\SO(3,2)$ representation $n$:
as $L^2<0$, we are only dealing with timelike intervals, and we
have a minimal time interval of length $n$ and then a continuum of
eigenvalues.

\subsection{The $\kappa$-Minkowski spacetime}
From the above discussion on the Snyder's picture of a non-commutative
spacetime, it should be clear that the fundamental physical and
mathematical inputs are simply the fact that space of momenta becomes
curved with a De Sitter geometry, the fact that, since this space is
best described as a coset $\SO(4,1)/\SO(3,1)$, there exist a global
symmetry algebra (and group) acting transitively on it, and the fact
that we can identify the spacetime coordinates, therefore the
configuration space of the theory, as a suitably defined ``translation
part'' of this symmetry algebra. However, in the definition of the
Snyder's picture, a choice was made in this last step, which was
natural from the group theoretic point of view but rather arbitrary
from the physical side: a choice of decomposition of the De Sitter
algebra into a Lorentz subalgebra and a ``translation part''; the
choice used was the usual Cartan decomposition, but this is not the
only possible. In other words,
there seems to be an ambiguity with respect to what one calls {\it
physical coordinates}. Indeed one could choose
another basis for the $x$'s and assume these are the true
spacetime coordinates, and there is nothing against this from the
mathematical point of view.

Now, is this ambiguity simply a formal one, an ambiguity in the
description of the same physics, or is it more than this? Are
different choices of coordinates equivalent from the physical point of
view, or is there a good argument for discarding some of them, thus
selecting what is {\it the} physical picture of non-commutative
spacetime of DSR? If they are all physically acceptable, what is their
respective meaning?

What seems to us a reasonable point of view on these issues is
that all the different possible choices of coordinates are
physically equivalent and to be allowed, but that they represent
different pictures of the same non-commutative spacetime as seen
by different observers; the point of view is then that of a {\it
relative non-commutative geometry}.

Let us look more closely to the concept of {\it relativity}.
Special relativity broke the classical notion of simultaneity: now
two observers going at different speeds would give different
accounts of events. By deforming the Poincar\'e group, DSR gives a
special role to the origin of the coordinate system used to
describe the events i.e to the position of the observer (and not
only its speed). Furthermore, by introducing a quantum spacetime
with non-commuting coordinates, defining an observer requires
providing the set of measurements she could do, i.e providing a
particular basis of the algebra. Now, the issue of choosing a
basis becomes: when one does an experiment measuring a coordinate,
or any other (geometric) observable, which operator are we
actually measuring?


\medskip

Let us study another special choice of coordinates called the
$\kappa$-Minkowski basis. It is defined, with respect to the
Snyder's basis, as
\be
X_0=x_0=t \qquad X_i=x_i+\f{a}{\hbar}K_i.
\ee

It is important to stress that this choice picks up a preferred time
direction, as compared to the Snyder's basis, in fact it corresponds
to a shift in the spatial coordinates but leaves intact the time
coordinate, and it is, as a
consequence, ``less covariant''. We feel that this \lq\lq less covariance" affects directly all the geometric properties we are about to discuss (and the coherent states constuction outlined below) in that they result in being tied to a given (class of) observer(s).

The reason why this choice of coordinates is special is most
easily understood by looking at the new commutators between
spacetime coordinates. These satisfy the following commutation
relations:
\be
[X_0,X_i]=-iaX_i \qquad [X_i,X_j]=0,
\label{comm}
\ee
so that now the coordinates satisfy Lie algebra commutation
relations forming thus a sub-algebra of the $\SO(4,1)$ algebra.
This subalgebra is interpreted as a non-commutative flat
spacetime, called the $\kappa$-Minkowski space. The deformation
parameter is defined as $\kappa=1/a$. When the length scale $a$ is
set to the Planck length, $\kappa$ thus becomes the Planck mass
(or energy).

In this picture the space coordinates are now commutative, so that
the quantum nature of the space operators is somewhat hidden,
however there is an intrinsic uncertainty when measuring both the
time coordinate and any of the space coordinates of an object, as
the corresponding operators do not commute. This uncertainty is
governed by the length scale $a$ and actually grows when getting
further from the origin. In the following, we will set $a=l_P$.

Concerning the other commutation relations, the Lorentz sector of the
$\SO(4,1)$ algebra is not
modified, as the only change with respect to the Snyder basis is in
the spacetime coordinates, and we still have the same commutation
relations between the $J$'s and the $K$'s.
Nevertheless the action of the boosts on the spacetime coordinates is now
modified:
$$
[K_i,X_0]=i\hbar\left(X_i-\f{a}{\hbar}K_i\right) \qquad
[K_i,X_j]=i\hbar\left(\delta_{ij}X_0-\f{a}{\hbar}\epsilon_{ijk}J_k\right).
$$
Of course also the relations between spacetime coordinate operators
and momentum operators (with a suitable choice of coordinates on De
Sitter space) is affected by the new definition of the former, the new
commutators being straightforwardly computable.

\medskip

Interestingly, the spectrum of the $X$'s operators (both time and
space) is now continuous! Then let us insist on the fact that the
$\kappa$-Minkowski basis actually depends on the choice of a time
direction which allows the split between space and time. A
possible interpretation is that the $\kappa$-Minkowski coordinates
are relative to an observer, when it describes physical phenomena
as happening in the usual classical continuous space (not
spacetime) with the time direction being defined along the
worldline of the observer. In this context, it appears that if one
wants to keep the notion of a classical space while having a
universal length scale $a$ existing for every observer, one needs
to deform the action of the Lorentz boosts and makes time fuzzy.

\medskip

The spacetime invariant length is defined again by the same operator
as before, but now its expression is modified and reads:
\be
L^2=X_0^2-\left(X_i-\f{a}{\hbar}K_i\right)\left(X_i-\f{a}{\hbar}K_i\right).
\ee

Now appears the difficulty of defining a natural notion of {\it
space distance} out of the space coordinate operators. Naturally,
if one assumes that the $X_i$ are the space coordinates, one would
introduce the following space length operator:
\be
\what{l}^2_\kappa\equiv X_iX_i=
\left(x_i+\f{a}{\hbar}K_i\right)\left(x_i+\f{a}{\hbar}K_i\right).
\ee
Let us stress that $\what{l}^2_\kappa\ne\what{l}^2$. Still
$\what{l}^2_\kappa$  is perfectly invariant under the $\SO(3)$
subgroup of space rotations and thus qualifies as a notion of
space distance. Nevertheless, this choice is not justified by the
expression for the invariant quadratic spacetime length and thus
is left with an unclear geometric interpretation. This results
from the unclear geometric interpretation of the new space
coordinates $X_i$ themselves (while the interpretation of the
Snyder coordinates as the non-commutative analogue of cartesian
coordinates was justified by the very expression for the spacetime
length operator). Notice that $\what{l}^2_\kappa$ has a continuous spectrum contrary
to $\what{l}^2$.

We then want to study the behavior of the newly defined space and
time distance operators under the action of the Lorentz subgroup
of symmetries, and in particular of the boosts, and the resulting
phenomenon of length contraction.

Nothing changes of course for the time distance operator, whose
definition remains the same, on the other hand it appears that there is a correction term in the
action of boosts on $X_1$:
$$
X_1^{(boosted)}=e^{i\eta
K_1}\left(x_1+\f{a}{\hbar}K_1\right)e^{-i\eta K_1}
=x_1^{(boosted)}+\f{a}{\hbar}K_1=
\cosh\eta x_1+\sinh\eta x_0 +\f{a}{\hbar}K_1
$$
\be
\Rightarrow
X_1^{(boosted)}=\cosh\eta X_1+\sinh\eta
X_0-\f{a}{\hbar}(\cosh\eta-1)K_1.
\ee
It is easy to check that here too $X_1$ and $X_1^{(boosted)}$ do
not commute:
$$
[X_1^{(boosted)}, X_1]=-ia\left(
\sinh\eta X_1 +(\cosh\eta-1)X_0
\right)\ne 0.
$$

Therefore again we have that if an observer sees a system at a
definite distance, i.e. in a given eigenstate of the $X_1$ operator, a
boosted observer will instead see a superposition of such states and
will not assign a definite distance to the same system, and again this is the key point that makes a Lorentz invariant discreteness of the quantum geometry possible.
The same difficulties we pointed out in the Snyder's case, in testing whether the usual length contraction rule is recovered for the mean values, are to be found also in this k-Minkowski basis, and a more complete treatment is left for future work.

We can again study what a point is in this new picture of
non-commutative spacetime, again using the notion of wave packets
to define it, and restricting the analysis to a 2-dimensional
slice of configuration space, to simplify it. Now, as
space coordinates do commute among themselves, the extension to
the full 4-dimensional picture is even more straightforward. A quantum
point, or, better, the state of an ideal system defining a given
location, will then be identified with a wave packet state for
which the spacetime coordinates would be as peaked as possible
around a given value of the spacetime coordinates $(X_0,X_i)$.

The commutation relations \Ref{comm} lead to a Heisenberg
uncertainty
relation of the type:
\be
\delta X_0 . \delta X_i \geq \f{l_P}{2} \, X_i,
\ee
so that the uncertainty on the variables $X_0,X_i$ increases with
$X_i$, i.e the further we are from the origin, or in other words
the further the object to be localized is from the observer, the
harder it is to localize it\footnotemark. Furthermore, we expect
that if an object is really well localized in time, say up to a
Planck scale resolution, i.e $\delta X_0=l_P$, then it would be
completely delocalized in space, i.e. with a delocalization equal
to its distance from the observer.

\footnotetext{Let us point out that this is consistent with the viewpoint
of the
$\kappa$-Minkowski coordinates being relative to a particular
observer standing at the origin: the further away an event is, the
harder it is to determine its precise time of occurrence.}

There is an easy way of constructing such wave packet states; let
us do a (non-linear) change of basis and introduce variables $y_i$
such that $X_i=a\exp(y_i/a)$; then we get the simple canonical
commutation relation:
\be
[X_0,y_i]=[x_0,y_i]=-ia^2=-il_P^2\sim -i\hbar,
\ee
which correspond to the classical bracket $\{x_0,y_i\}=1$. We
would like to introduce localized states which minimize the
uncertainty on $x_0$ and the coordinates $y_i$. To this purpose,
we can use the usual wave packet states as built for a harmonic
oscillator on the variables $x_0,y_i$ in order to minimize the
uncertainties $\delta x_0\sim\delta y_i\sim \sqrt{\hbar}=l_P$.

Such states have the simple gaussian form:
$$
f_{(x_0^0,\vec{y}^0)}(y_i)\propto
e^{i\f{x_0^0y_1}{a^2}}e^{i\f{x_0^0y_2}{a^2}}e^{i\f{x_0^0y_3}{a^2}}e^{-\f{|\vec{y}-\vec{y^0}|^2}{a^2}}.
$$
In the $X_i$ coordinates, these states have a more complicated
form.

However, these are not good localized states with respect to the
$\kappa$-Minkowski coordinates, as can be seen by simply looking
again at the uncertainty relations: the uncertainty on the $X_i$'s
is:
\be
\delta X \sim e^{\f{y}{l_P}}\delta y \sim \f{X}{l_P}\delta y \sim X.
\ee
Indeed we see that such states are completely delocalized in
space.

However, a simple modification of the above construction solves
the problem: we introduce an additional length scale $l$ for
semi-classical physics, such that $l >> l_P$ but still much
smaller that the resolution of any achievable direct measurement
of localization (for example, the best resolution in time ever
been achieved, and very recently, is of $10^{-16}$ seconds,
hundreds of attoseconds, while the Planck time is of the order
of $10^{-40}$ seconds, so there is plenty of room in between!).
Now, let us build coherent states for the conjugated pair of
canonical variables $(x_0\alpha, y_i/\alpha)$ with $\alpha=l/l_P$.
Then the magnitude of the uncertainties are:
\be
\delta x_0=l \quad\textrm{ and }\quad
\delta y_i=\f{1}{\alpha} \,\Rightarrow\,
\delta X_i=\f{1}{\alpha}{X_i}=\f{l_P}{l}X_i.
\ee

Choosing $\alpha\ll 1$ such that $l_P\ll l \ll {\cal L}$, where
${\cal L}$ corresponds to our length scale (the scale of
resolution of our best measurement devices) where all looks smooth
and classical, we can consider such the  as semi-classical states
where spacetime points are  well localized, or simple {\it
localized states}\footnotemark. \footnotetext{Let us point out that such a setting is one of the current
effective framework for extracting semi-classical information out of a
Loop
Quantum Gravity state \cite{AMU}. Actually, the introduction of an additional quantity characterizing the scale of observations at which the effective description applies is a general feature of any approximation scheme.}

\medskip

A complete analysis of the transformation properties under boosts of
these wave packets is left then for future work. The issue is again how
localized states with respect to the boosted coordinate operators
differ from the one constructed with respect to the un-boosted ones.
Let us finish by pointing out that, interpreting the $\kappa$-Minkowski basis as depending on an observer and its time direction, then in order to build coherent states for a boosted observer we would need to change basis by changing the $x_0$ direction in the algebra and then re-do the whole work to construct new coherent states which would be completely different than the ones constructed for the original observer. From this point of view, it is clear that the description of events by a boosted observer will be different from the description provided by the original observer. This again highlights the lack of covariance of this basis as compared to the Snyder's basis, but simply because it is attached to a particular observer.

\section{Comparison of results and hints for Loop Quantum Gravity}

Here we should add a few comments on the
results obtained, and a comparison between what we got in DSR case
and the toy model inspired by LQG, stressing similarities in the
overall picture, in spite of differences in the details.

Let us comment briefly on the results obtained in the toy
loop-inspired model and in the DSR framework. The basic point that is
in common to both cases is a fundamental Lorentz invariance of the
setting. At the same time in both cases it is possible to give a
simple definition of geometric distance operators giving rise to a
discrete picture of quantum geometry. The apparent contradiction
between a continuous group of symmetries and this discreteness is
solved thanks to the unitary representation of this group on the
Hilbert space of states of the theory and to the non-commutativity
between geometric operators and their Lorentz-transformed
counterparts.
Again, this is true, and realized similarly, both in the Loop-inspired
toy model and in DSR. In the first framework, moreover, due to its
extreme simplicity, it is possible to check explicitly that one
recovers the usual transformation laws for space and time distances,
i.e. the usual Lorentz contraction or dilatation, for the mean values
of the relevant operators. In the DSR case the more involved details
of the framework prevents from doing so.
However, also in this case we would expect the usual laws to be
recovered for mean values. The reason is that the DSR framework is
usually interpreted as an effective description of the flat limit of
Quantum Gravity; in this limit, i.e. if the quantum state of geometry
under consideration is that describing flat space, we should have that
the Lorentz symmetry extends from a local symmetry of the theory to a
global isometry of the geometry, in other words to a global symmetry
of the state, either exactly or in a macroscopic limit. Standing the
quantum nature of the problem and of the setting, and thus the results
found above about the behavior under boosts of the geometric
operators, we expect this to be manifest exactly in the classical-like
behavior of the mean values of these geometric operators.
One can also turn this logic around and use the properties of
geometric operators in the given state to gain information about the
state itself. In other words, the presence
of such a global symmetry, made clear by the classical-looking
transformation properties of the expectation values of distance
operators, can be used to characterize the
state under consideration as that corresponding to flat space (of
course also a translation symmetry has to be present in addition to
the Lorentz one). This is a hint of a new possible approach to
defining a 'Minkowski vacuum state' in Loop Quantum Gravity (LQG), inspired
by the analysis presented in this paper.
Let us make this more explicit.

In \cite{simone}, the authors defined Lorentz transformation in LQG and studied
the properties of the Lorentz boosted area operator. Here we propose the idea of applying
the same framework to the volume operator. Lorentz boosts are generated by the Hamiltonian
constraint ${\cal H}$ with a special choice of lapse function $N$ depending on the boost rapidity $\eta$.
The (Poisson) algebra of commutation between the volume (operator) ${\cal V}$ and the Hamiltonian ${\cal H}$ is
particularly interesting. Roughly it goes as:
\bes
{\cal H} &=& \{S_{CS},{\cal V}\}, \nn\\
{\cal C} &=& \{{\cal H},{\cal V}\}, \nn\\
\f{2}{3}{\cal V} &=& \{{\cal C},{\cal V}\},
\ees
where $S_{CS}$ is the Chern-Simons action and ${\cal C}$ the operator implementing changes of Immirzi parameter in LQG and thus generating scale transformation on the kinematical geometrical operators (like the volume).
Now, under a small boost, the volume will get shifted to:
\be
{\cal V}_{\eta}={\cal V}+\{N_\eta{\cal H}, {\cal V}\}={\cal V}+N_\eta{\cal C}.
\ee
It is obvious that ${\cal V}_{\eta}$ and ${\cal V}$ do not commute. Moreover their commutators is proportional
to the volume itself:
\be
\{{\cal V}_{\eta},{\cal V}\}\propto{\cal V},
\ee
which implies that the volume and boosted volume become more and more "different" when the volume grows i.e their eigenstates are different.
Now requiring the contraction law of the volume under boosts in the flat Minkowski background, $V_\eta=V/\cosh\eta$,
a (localized)  Minkowski state $|\psi\ra$ should satisfy the following criteria:
\be
\la\psi|{\cal V}_\eta|\psi\ra\approx\f{\la\psi|{\cal V}|\psi\ra}{\cosh\eta}
\,\Rightarrow\,
\la\psi|N_\eta{\cal C}|\psi\ra\approx \f{-\eta^2}{2}\la\psi|{\cal V}|\psi\ra,
\ee
for (macroscopic) space regions, when $\la\psi|{\cal V}|\psi\ra$ is large and $\eta$ small. Further analysis of
the operator ${\cal C}$ is required to check how meaningful this proposed criterium is.

\section*{Conclusion}

The main goal of this paper was to show in which sense the
discreteness of a quantum spacetime geometry originating from Quantum
Gravity may still be compatible with Lorentz invariance; this is
somewhat counterintuitive, as almost everything when it comes to
Quantum Gravity, but we have shown how this may be possible and easily
realized if one takes seriously the quantum nature of geometric
measurements in this context.

In order to achieve this result, we studied a simplified model
of a quantum flat geometry, directly inspired by 3-dimensional Loop
Quantum Gravity, where it was possible to define what time and space
distance measurements were and to carry through all the relevant
(easy) calculations. We think this simple model can be of inspiration
also for work in the complete Loop Quantum Gravity setting.

We found that the compatibility of quantum discreteness of geometric
spectra and continuous Lorentz invariance is possible due to the
unitary action of the Lorentz boost operators  on quantum states and
distance operators, and a non-commutativity of these and their boosted
counterparts. This results in the fact that the state of a localized
system for a given observer turns into a de-localized one for another
observer boosted with respect to the first.
Our result then confirms, in this simplified context, but in full
detail, the argument for resolving the apparent contrast between discrete
quantum geometry and Lorentz invariance presented in \cite{simone}.

We think it is also of interest the fact that a simple procedure
for mapping $\SU(2)$ (Riemannian) quantum geometry to the $\SU(1,1)$
(Lorentzian) one, introduced by 't Hooft in \cite{thooft} turned out
to be useful in our analysis; this may hint to a closer than expected
similarity between the Loop Quantum Gravity/Spin Foam approach and his
polygon-based  discrete framework.

In the second part of the paper, we turned our attention to Doubly (or
Deformed) Special Relativity theories, considered as effective models
of full quantum gravity. In this models Lorentz invariance is present
but somehow hidden, due to a modified action of the Lorentz group, and
most important the picture of quantum geometry that lies behind them
is unclear. We tried to analyze this important aspect using similar
tools as those which are customary in Loop Quantum Gravity and Spin
Foam Models, basing our work on the geometric rather than on the quantum
algebraic structure of DSR, in both the Snyder and the
$\kappa$-Minkowski bases.

We gave a simple geometric definition of time interval and space
distance measurement operators in DSR, and analyzed the spectral
properties of these and of other geometric operators; this needed a
first study of the Hilbert space structure of DSR models.
We then analyzed the behavior under boosts of these operators and of
their spectra, showing a nice parallelism with the structure and
properties of the loop inspired toy model discussed previously.
Again it is the quantum nature of geometric operators that makes the
co-existence of discrete quantum geometry and continuous Lorentz
invariance possible.
We also discussed wave packet states, as a sensible definition of
quantum spacetime points in a non-commutative context.

In this way, we think we achieved a better, although far from
complete, understanding of the picture of quantum geometry resulting
from DSR theories, and made the comparison between the full Loop
Quantum Gravity/Spin Foam theory and its DSR effective description
a bit easier.
Also, we think the analysis presented may give useful hints on the
problem of defining and characterizing a quantum state corresponding
to flat space in a full theory of Quantum gravity, e.g. Loop Quantum Gravity.

We are well aware that this can be at best a very first step towards a
complete understanding of the issues dealt with in this paper, but we
feel it was a {\it necessary} first step, and a valuable one.

\section*{Acknowledgements}

We would like to warmly thank Laurent Freidel, Florian Girelli,
Carlo Rovelli and Lee Smolin for useful comments and discussions; D.O. thanks also Jurek Kowalski-Glikman, Gianluca Mandanici and Giovanni Amelino-Camelia for many helpful discussions on DSR during the 40th Karpacz Winter School on 'Quantum Gravity Phenomenology'.

\appendix

\section{Length contraction in Special Relativity}

Let us consider a ruler of length $l$ in its  referential at rest. Let us
consider an
observer moving at a speed $v$ (in the direction defined by the axis of
the ruler). Then
the two referentials are related by a Lorentz transformation $\Lambda$ and
the length
$l'$ seen by the moving observer is defined as:
\be
\mat{c}{t' \\x'}=\Lambda\mat{c}{t \\x}=
\mat{cc}{\gamma & \gamma v \\ \gamma v & \gamma}\mat{c}{t \\x}
\qquad \Rightarrow \qquad
\mat{c}{0 \\l'}=\mat{cc}{\gamma & \gamma v \\ \gamma v & \gamma}\mat{c}{t
\\l},
\ee
with $\gamma=1/\sqrt{1+v^2}=\cosh\eta$, $\eta$ being the boost rapidity.
Then solving this equation, one gets the usual result: $l'=l/\gamma$.

Let us keep in mind that we are dealing with the distance between two spacetime points, but between the two worldlines of the ends of the ruler. Each observer is going to give a different values of such a distance. The Lorentz invariant notion is the concept of the {\it maximal} distance between the two lines, which is indeed given by the rest ruler length $L$ (the minimal positive distance is of course 0, since there always exist a null ray going from one worldline to the other).

\medskip

Let us compare this result with the true length measurement done through the photon time-of-flight
experiment. For an observer ${\cal O}_1$ moving towards the other end of the ruler, the photon flight
will be shorter than for the observer at rest ${\cal O}$. It is easy to check that:
$$
T_1=T\f{1}{\gamma(1+v)}=Te^{-\eta}\,<\,T, \quad \eta\ge0.
$$
One should note that the contraction factor is not $\gamma=\cosh\eta$ anymore.
Now if an observer ${\cal O}_2$ moves away from the ruler, then the photon flight will be longer than for ${\cal O}$:
$$
T_2=Te^{+\eta}\,>\,T, \quad \eta\ge0.
$$

\section{Some facts about Snyder's DSR}

To complete our presentation of DSR, let us add a few facts. In the Snyder basis, as the spectrum of some
coordinate operators is discrete, one can not represent the momentum operators $p_\mu$ as derivation operators
$\pp/\pp x_\mu$. On the other hand, one can use the $p$-polarisation and still represent  the $x_\mu$
operators as derivations with respect to the momenta \cite{snyder}:
\be
\what{x_i}=i\hbar\left(
\f{\pp}{\pp p_i}+\left(\f{a}{\hbar}\right)^2p_i\left(
p_1\f{\pp}{\pp p_1}+p_2\f{\pp}{\pp p_2}+p_3\f{\pp}{\pp p_3}+p_t\f{\pp}{\pp
p_t}
\right)
\right),
\ee
\be
\what{t}=i\hbar\left(
\f{\pp}{\pp p_t}-\left(\f{a}{\hbar}\right)^2p_t\left(
p_1\f{\pp}{\pp p_1}+p_2\f{\pp}{\pp p_2}+p_3\f{\pp}{\pp p_3}+p_t\f{\pp}{\pp
p_t}
\right)
\right).
\ee
These operators are then Hermitian with respect to the deformed measure:
\be
d^4\tau=\f{\hbar \,dp_1\,dp_2\,dp_3\,dp_t}{ac\left(
p_1^2+p_2^2+p_3^2-p_t^2+\left(\f{\hbar}{a}\right)^2\right)},
\ee
which is to be used in quantum mechanics (or field theory) computations. We see that it contains a singularity, or equivalently a resonance, at the Planck mass $m^2=(\hbar/a)^2$, which is a signature of some quantum gravity effects.

Let us underline that the previous expression were given in \cite{snyder} for the DeSitter case.
In the DSR theory defined as an AntiDeSitter lattice, the correction would change sign and we would
have a $-(\hbar/a)^2$ term in the denominator of the measure: upon restriction to purely timelike momenta, there is no divergence in the measure anymore at $p=0$.

\section{About representations of $\SO(2,1)$, $\SO(3,1)$ and $\SO(4,1)$}

The Lorentz groups in 3d,4d and 5d are respectively $\SO(2,1)$, $\SO(3,1)$
and $\SO(4,1)$. We
need to study their representation theory inorder to extract the
properties of the length
operator of Doubly Special Relativity.Let us choose as signature $(-++..)$
so that spacelike
curves have a positive squared length.

\medskip

$\SO(2,1)$ has 3 generators $J_{01,02,12}$ (two boosts and one rotation).
It has one Casimir operator:
$$
C(\SO(2,1))=J_{12}^2-J_{01}^2-J^2_{02}.
$$
Its (principal) unitary representations are of two kinds:
\begin{itemize}
\item the continuous series labeled by a real number $s>0$.
For them, $C(\SO(2,1))=-(s^2+1/4)$.
\item the discrete series labeled by an integer $j>1$ and a sign
$\epsilon$.
Then $C(\SO(2,1))=j(j-1)$.
\end{itemize}

\medskip

$\SO(3,1)$ has 6 generators $J_{IJ}$ ($I,J=0,..,3$). Its has two
Casimir operators: $C(\SO(3,1))=J_{IJ}^2-J_{0I}^2$ and
$\wtl{C}(\SO(3,1))=\epsilon^{IJKL}J_{IJ}J_{KL}$. Its (principal)
unitary representations are labeled by a couple
$(n\in\N,\rho\in\R^+)$ and their Casimir are:
$C(\SO(3,1))=n^2-\rho^2-1$ and $\wtl{C}(\SO(3,1))=2n\rho$. We call
{\it simple representations} these ones that have a vector
invariant under the $\SO(2,1)$ subgroup (leaving the third
direction $x_3$ invariant). Obviously, the second Casimir then
vanishes, $n\rho=0$, and we have two series of representations:
$(n,0)$ and $(0,\rho)$. Let us also point out that a
representation $(n,\rho)$ can be decomposesd into $\SO(3)$
representations and that the relevant spins are $j\ge n$.

\medskip

$\SO(4,1)$ has 10 generators $J_{IJ}$ ($I,J=0,..,4$). It also has two
Casimir
operators: $C(\SO(4,1))=J_{IJ}^2-J_{0I}^2$ and
$\wtl{C}(\SO(4,1))=-W_0^2+W_IW_I$ with $W_I=\epsilon^{IABCD}J_{AB}J_{CD}$.
Let us notice that $W_4=C(\SO(3,1))$.
Simple representations are now defined as having a $\SO(3,1)$ invariant
vector.
It is straightforward to check that this is equivalent to
$\wtl{C}(\SO(4,1))=W^2=0$, so that we are also left with two series of
representations $(N,0)$ and $(0,R)$.


\end{document}